\documentclass[aps,prl,twocolumn,showpacs]{revtex4}
\usepackage{amsmath,epsfig,amsfonts,epsfig,graphicx,
}

\begin{document}
\title{Feshbach Resonance Management of Bose-Einstein Condensates in Optical Lattices}

\author{Mason A. Porter}
\affiliation{Department of Physics and Center for the Physics of
Information, California Institute of Technology, Pasadena, CA USA 91125}

\author{Marina Chugunova}
\affiliation{Department of Mathematics \& Statistics, McMaster
University, Hamilton, Ontario, Canada L8S 4K1}

\author{Dmitry E. Pelinovsky}
\affiliation{Department of Mathematics \& Statistics, McMaster
University, Hamilton, Ontario, Canada L8S 4K1}

\begin{abstract}
We analyze gap solitons in trapped Bose-Einstein condensates (BECs) in optical lattice potentials under Feshbach resonance management.  Starting with an averaged Gross-Pitaevsky (GP) equation with a periodic potential, we employ an envelope wave approximation to derive coupled-mode equations describing the slow BEC dynamics in the first spectral gap of the optical lattice. We construct exact analytical formulas describing gap soliton solutions and examine their spectral stability using the Chebyshev interpolation method.  We show that these gap solitons are unstable far from the threshold of local bifurcation and that the instability results in the distortion of their shape.  We also predict the threshold of the power of gap solitons near the local bifurcation limit.
\end{abstract}

\date{\today}

\maketitle

\vspace{2mm}

\section{Introduction}

At sufficiently low temperatures, particles in a dilute boson gas 
can condense in the ground state, forming a Bose-Einstein condensate 
(BEC) \cite{stringari}.  Under the typical confining conditions of 
experimental settings, BECs are inhomogeneous and the number of 
condensed atoms ($N$) ranges from several thousand (or less) to 
several million (or more). The magnetic traps that confine them are 
usually approximated by harmonic potentials.  There are two 
characteristic length scales: the harmonic oscillator length $a_{ho} 
= \sqrt{\hbar/(m\omega_{ho})}$ [which is on the order of a few 
microns], where $\omega_{ho}=(\omega_x \omega_y \omega_z)^{1/3}$ is 
the geometric mean of the trapping frequencies, and the mean healing 
length $\chi=1/\sqrt{8\pi |a| \bar{n}}$ [which is on the order of a 
micron], where $\bar{n}$ is the mean particle density and $a$, the 
(two-body) $s$-wave scattering length, is determined by the atomic 
species of the condensate.  Interactions between atoms are repulsive 
when $a > 0$ and attractive when $a < 0$.  For a dilute ideal gas, 
$a \approx 0$.

If considering only two-body, mean-field interactions, a dilute
Bose-Einstein gas can be modeled using a cubic nonlinear 
Schr\"odinger equation (NLS) with an external potential; this is 
also known as the Gross-Pitaevsky (GP) equation.  BECs are modeled 
in the quasi-one-dimensional (quasi-1D) regime when the transverse 
dimensions of the condensate are on the order of its healing length 
and its longitudinal dimension is much larger than its transverse 
ones \cite{kohler}. The GP equation for the condensate wavefunction 
$\psi(x,t)$ takes the form
\begin{equation}
    i\hbar \psi_t = -\frac{\hbar^2}{2m} \psi_{xx} + V(x) \psi + g |\psi|^2\psi\,, 
    \label{nls3}
\end{equation}
where $|\psi|^2$ is the number density, $V(x)$ is the external
trapping potential, $g = [4\pi\hbar^2 a/m][1 +
\mathcal{O}(\zeta^2)]$ is proportional to the two-body scattering
length, and $\zeta = \sqrt{|\psi|^2 |a|^3}$ is the dilute gas
parameter \cite{stringari,kohler}.

Experimentally realizable potentials $V(x)$ include harmonic traps, 
quartic double-well traps, optical lattices and superlattices, and 
superpositions of lattices or superlattices with harmonic traps. The 
existence of quasi-1D (``cigar-shaped'') BECs motivates the study of 
lower dimensional models such as Eq.~(\ref{nls3}).  We focus here 
on the case of spatially periodic potentials without a confining 
trap along the dimension of the lattice, as that is of particular 
theoretical and experimental interest.  For example, such potentials 
have been used to study Josephson effects \cite{anderson}, squeezed 
states \cite{squeeze}, Landau-Zener tunneling and Bloch oscillations 
\cite{morsch}, period-multiplied wavefunctions 
\cite{mapbecprl,pethick2}, and the transition between superfluidity 
and Mott insulation at both the classical \cite{smerzi} and quantum 
\cite{mott} levels.  Moreover, with each lattice site occupied by 
one alkali atom in its ground state, a BEC in an optical lattice 
shows promise as a register in a quantum computer \cite{voll}.

The properties of BECs---including their shape, collective
excitations, statistical fluctuations, and the formation and
dynamics of their solitons and vortices---are determined by the
strength and sign of their two-body atomic interactions $a$.  This
scattering length, and hence the coefficient of the nonlinearity
in the GP equation, can be adjusted in both sign and magnitude
(over a large range) by minutely adjusting a magnetic field in the
vicinity of a so-called ``Feshbach resonance" \cite{fesh,inouye}.

A Feshbach resonance is an enhancement in the scattering amplitude
of a particle incident on a target when the energy of the former
is approximately that needed to create a quasibound state of the
two-particle system.  If a pair of ultracold atoms has a molecular
bound state near zero energy, then during collisions they stick
together for a little while as they undergo a Feshbach resonance.
While few molecules have bound states at such energies, one can
adjust the relative energies of atoms and molecules with different
magnetic moments by applying a magnetic field.  With such ``Zeeman
tuning", one can move the atomic energy from just above the
resonance to just below it, so that the scattering length diverges
and changes sign from positive to negative across the resonance.

As a result of the control this procedure gives over condensate 
properties, the manipulation of ultracold atoms using Feshbach 
resonances has become among the most active research areas in 
experimental atomic physics.  Feshbach resonances have provided a 
key for creating molecular BECs, generating solitons and 
atom-molecule coherence, stabilizing or destabilizing BECs, and 
creating novel Fermi liquids \cite{feshreview,klep,donley2,regal1}.  
For example, it was recently shown that near a Feshbach resonance, a 
quantum phase transition occurs between a regime with both atomic 
and molecular condensates and one with only molecular condensates 
\cite{romans}.  As pointed out in Ref.~\cite{stoof2}, this 
transition should be much easier to observe for condensates loaded 
into optical lattice potentials.

In Feshbach resonance management, which was motivated by similar 
techniques in fiber optics \cite{kurtzke}, the BEC scattering length 
is varied periodically in time.  This yields dynamically interesting 
soliton solutions, such as breathers \cite{FRM}. Very recently, 
there has been some theoretical work concerning Feshbach resonances 
in BECs in optical lattices \cite{tsoy,feshol,matu2}. This situation 
is also the subject of current experimental investigations 
\cite{esslinger}.

In the present paper, we use an averaged GP equation \cite{hamavg,zharn} to examine gap solitons in BECs trapped in optical lattices and under the influence of Feshbach resonance management. Using an envelope-wave approximation, we derive coupled-mode equations describing the slow dynamics of gap solitons.  We provide an analytical construction of gap soliton solutions and examine their spectral stability with numerical computations of eigenvalues. We then describe the time-evolution of gap solitons with the averaged and original GP equations.  Finally, we summarize our results.

\section{The averaged Gross-Pitaevsky equation}\label{avg}

We consider the non-dimensional GP equation for trapped BECs under Feshbach resonance management,
\begin{equation}
    \label{GP} i\psi_t = -\psi_{xx} + V(x)\psi + g(t) |\psi|^2 \psi\,,
\end{equation}
where the normalized independent variables are 
\begin{equation*}
	\tilde{x} = \frac{\sqrt{2m} x}{\hbar}\,, \qquad \tilde{t} = \frac{t}{\hbar}\,,
\end{equation*}
and the tildes have been dropped from Eq.~(\ref{GP}).  The nonlinear coefficient can be written \cite{FRM}
\begin{equation}
    g(t) = \gamma_0 + \frac{1}{\varepsilon}\gamma\left(\frac{t}{\varepsilon}\right)\,, \label{timedep}
\end{equation}
and the potential for optical lattices is \cite{pelsukhkiv}
\begin{equation}
    	V(x) = 2 \epsilon V_0 \cos(\omega x)\,. \label{four}
\end{equation}
Here, $\gamma_0$ is the mean value of the nonlinearity coefficient; $\gamma(\tau)$, with $\tau = t/\varepsilon$, is a mean-zero periodic function with unit period; $\varepsilon \ll 1$ is a small parameter 
describing the strength of the Feshbach resonance management; $\omega$ is the wavenumber of the optical lattice; and $\epsilon V_0$ is a small parameter describing the strength (amplitude) of the lattice. Because (\ref{timedep}) changes rapidly in time (i.e., the management is strong), it is reasonable use a model without dissipation \cite{esslinger}. We note that the two small parameters $\epsilon$ and $\varepsilon$ in Eq.~(\ref{GP}) are due to two different physical sources and can be uncorrelated.  

One can exert very precise control over optical lattice strengths and wavenumbers experimentally \cite{olstandard}.  In particular, both strong and weak lattices can be implemented; we will consider a weak optical lattice, so that $\epsilon$ in (\ref{four}) is small.  In experiments, the scattering length $a$ can also be adjusted either nonadiabatically (strong nonlinearity management), which is the situation discussed in the present work, or adiabatically (weak nonlinearity management).  In general, different dynamics can occur depending on how rapidly $a$ is adjusted, as discussed, e.g., in \cite{cubi}.  There have already been some experiments in which Feshbach resonances are applied to BECs in the presence of optical lattices \cite{esslinger} and others are being planned by multiple experimental groups.  As mentioned above, a conservative model is appropriate for strong nonlinearity management.  For weak nonlinearity management, however, the GP equation (\ref{GP}) needs to be augmented by dissipation terms, as three-body recombination leads to experimental losses \cite{fesh,esslinger}.  Therefore, we will only consider strong management, so that $\varepsilon$ in (\ref{timedep}) is small.

The small parameter $\varepsilon$ can be used to simplify the
time-periodic GP equation (\ref{GP}) with an averaging method. Using
the time-averaging procedure from \cite{hamavg}, one can look for an 
asymptotic solution to the GP equation (\ref{GP}) of the form
\begin{equation}
    \label{transformation} \psi(x,t) = e^{i \varepsilon
    \gamma_{-1}(\tau) |u|^2(x,t)} \; \left[ u(x,t) + {\rm
    O}(\varepsilon) \right]\,,
\end{equation}
where $u$ is the complex-valued amplitude and $\gamma_{-1}(\tau)$ is
the mean-zero anti-derivative of $\gamma(\tau)$. Within a regular averaging procedure 
(see the review in \cite{zharn}), we obtain the averaged GP equation,
\begin{align}
    i u_t &= -u_{xx} + 2 \epsilon V_0 \cos(\omega x) u + \gamma_0
|u|^2 u \notag \\ &\quad - \gamma_1^2\left[((|u|^2)_x)^2 +
2|u|^2(|u|^2)_{xx}\right] u\,, \label{avg2}
\end{align}
where $\gamma_1$ is the standard deviation of the nonlinearity 
coefficient. The model (\ref{avg2}) provides a starting point for 
our analysis of gap solitons in BECs in optical lattices under 
Feshbach resonance management. Without such management ($\gamma_1 = 
0$), gap solitons of the GP equation (\ref{avg2}) were studied in 
\cite{pelsukhkiv} using Floquet theory, multiple-scale expansions, 
beyond-all-orders theory, and Evans-function computations of 
eigenvalues.  We will mainly consider the opposite case when $\gamma_1 \neq 0$ but $\gamma_0 = 0$.

\section{Coupled-mode Equations} \label{cmes}

We are interested in modeling gap solitons supported by 
subharmonic resonances between the periodic potential and 
spatio-temporal solutions of the averaged GP equation (\ref{avg2}). 
To simplify the model, we use the second small parameter $\epsilon$ 
and obtain coupled-mode equations, which averages the spatially 
periodic nonlinear equation (\ref{avg2}) near a spectral gap of its 
associated linearization,
\begin{equation}
    \label{periodic-potential} i u_t = - u_{xx} + 2 \epsilon V_0
    \cos(\omega x) u\,.
\end{equation}
In the limit of small $\epsilon$, the spectral gaps all become narrow 
and the first spectral gap occurs at first order in $\epsilon$. 
Using the space-averaging technique of Ref.~\cite{agueev}, one can 
look for an asymptotic solution of the averaged GP equation 
(\ref{avg2}) in the two-wave form
\begin{align}
    \label{two-wave} u(x,t) &= \sqrt{\epsilon} \left(
    A(X,T) e^{i\omega_0 x - i\omega_0^2t}  \right. \notag \\
    &\quad \left. + B(X,T)e^{-i\omega_0x-i\omega_0^2t} + {\rm O}(\epsilon) \right)\,,
\end{align}
where $A$ and $B$ are complex-valued amplitudes, $X = \epsilon x$ 
and $T = \epsilon t$ are slow variables, and $\omega = 2\omega_0$. 
This wavenumber ratio indicates that we are studying $2\!:\!1$ 
subharmonic resonances [e.g. the first spectral gap of equation 
(\ref{periodic-potential})]. Using the regular asymptotic procedure 
from \cite{agueev}, we obtain a system of coupled-mode equations
\begin{align}
    i(A_T + \omega A_X) &= V_0 B + \gamma_0(|A|^2 + 2|B|^2)A
\notag \\ &\quad + 2\gamma_1^2\omega^2(2|A|^2 + |B|^2)|B|^2A\,, \notag \\
    i(B_T - \omega B_X) &= V_0A + \gamma_0(2|A|^2 + |B|^2)B
\notag \\ &\quad + 2\gamma_1^2\omega^2(|A|^2 + 2|B|^2)|A|^2B\,. \label{cme}
\end{align}
The first equation governs the left-propagating wave $A$, whereas the second equation governs the
right-propagating wave $B$. The two waves interact with a cubic cross-phase modulation from the mean value of the scattering length and with a quintic cross-phase modulation from the standard deviation
of the scattering length. The latter effect represents the main contribution of Feshbach resonance management in trapped BECs.

The system of coupled-mode equations (\ref{cme}) is Hamiltonian, with symmetric potential energy
\begin{align}
    W(A,B) &= V_0 \left( \bar{A} B + A \bar{B} \right) \notag \\ &\quad +
    \frac{\gamma_0}{2} \left( |A|^4 + 4 |A|^2 |B|^2 + |B|^4 \right) \notag \\ &\quad+ 
    2\gamma_1^2 \omega^2 |A|^2 |B|^2 (|A|^2 + |B|^2)\,.
    \label{W}
\end{align}
Additionally, Eq.~(\ref{W}) satisfies the assumption on symmetric 
potential functions used recently for analyzing the existence and 
stability of gap solitons \cite{chugpel}. While the previous work 
concerned gap solitons in coupled-mode equations with cubic 
nonlinearity, we will focus on the new effects that arise from the 
quintic nonlinear terms.  These effects correspond to mean-zero 
Feshbach resonance management, which affects the propagation of gap 
solitons in optical lattices. We can see that the last term of $W$ 
in (\ref{W}) is positive definite, similar to the second term with 
$\gamma_0 > 0$.  Therefore, Feshbach resonance management leads to 
defocusing effects on the propagation of gap solitons in periodic 
potentials.  The defocusing role of Feshbach resonance management 
was studied recently in \cite{kevstefpel} in the context of blow-up 
arrest in multi-dimensional GP equations.

One determines the linear spectrum of the coupled-mode  equations
(\ref{cme}) from the linearized system in Fourier form, $(A,B)
\sim e^{i K X - i \Omega T}$, where $\Omega = \pm \sqrt{V_0^2 +
\omega^2 K^2}$.  The spectral gap exists for $|\Omega| < |V_0|$
and corresponds to the first spectral gap associated with the periodic
potential in Eq.~(\ref{periodic-potential}). The lower (upper) spectral
band of the coupled-mode equations (\ref{cme}) for $\Omega <
-|V_0|$ (for $\Omega > |V_0|$) corresponds to the first (second)
spectral band of the periodic potential in (\ref{periodic-potential}).

The coupled-mode system (\ref{cme}) can be reduced to a nonlinear 
Schr\"{o}dinger (NLS) equation near the band edges of the linear 
spectrum. Using the asymptotic representation
\begin{align}
    A &= \sqrt{\mu} e^{\pm i |V_0| T} \left[ W(\xi,\zeta) \pm
    \frac{i \mu \omega}{2 V_0} W_{\xi} + {\rm O}    (\mu^2) \right]\,, \\
    B &= \sqrt{\mu} e^{\pm i |V_0| T} \left[ \mp W(\xi,\zeta) + \frac{i
    \mu \omega}{2 V_0} W_{\xi} + {\rm O}(\mu^2) \right]\,,
\end{align}
where $\xi = \mu X$, $\zeta = \mu^2 T$, and $\mu$ is a small
parameter for the distance of $\Omega$ from the band edges $\pm 
|V_0|$, one can reduce the coupled-mode equations (\ref{cme}) with 
$\gamma_0 = 0$ to the quintic NLS equation:
\begin{equation}
    \label{quantic-NLS} i W_{\zeta} = \pm \frac{\omega^2}{2 V_0} W_{\xi
    \xi} + 6 \gamma_1^2 \omega^2 |W|^4 W\,.
\end{equation}
The quintic NLS equation (\ref{quantic-NLS}) is focusing near the
lower spectral band with $\Omega = -|V_0|$ and is defocusing near
the upper spectral band with $\Omega = |V_0|$. Therefore, the gap
soliton solutions bifurcate from the lower spectral band via a local 
(small-amplitude) bifurcation.  They terminate at the upper spectral 
band via a nonlocal (large-amplitude) bifurcation, similar to gap 
solitons in the GP equation with a periodic potential \cite{pelsukhkiv}. 

The derivation of the quintic NLS equation (\ref{quantic-NLS}) 
confirms the predictions of the recent paper \cite{malfesh05} on the 
power threshold for one-dimensional gap solitons in the case 
$\gamma_0 = 0$ and $\gamma_1 \neq 0$. The existence of the power 
threshold near the local bifurcation limit was computed numerically in Ref.~\cite{malfesh05}. Because the quintic NLS equation exhibits a similar power threshold for NLS solitons 
(see, e.g., \cite{pelkivafan}), the numerical fact is now confirmed 
from the perspective of asymptotic theory. We note that the 
coupled-mode equations (\ref{cme}) with $\gamma_0 \neq 0$ reduce to 
the cubic NLS equation, which does not exhibit the power threshold 
for NLS solitons.

\section{Gap Solitons} \label{gap}

We simplify the construction of exact gap soliton solutions to the
coupled-mode equations (\ref{cme}) by normalizing $V_0 = -1$,
$\omega = 1$ (a standard scaling transformation can be employed for this purpose) and defining
$\sigma^2 = 2 \gamma_1^2 \omega^2$.  We construct gap
soliton solutions by separating variables into time-periodic and
spatially localized solutions of the coupled-mode equations (\ref{cme}):
$$
    A(X,T) = a(X) e^{- i \Omega T}\,, \quad B(X,T) = b(X) e^{-i \Omega T}\,.
$$
Because of the symmetry in the potential function, $W(A,B) = W(B,A)$, the
gap soliton solutions satisfy the constraint $b = \bar{a}$
\cite{chugpel}, so that $a(X)$ solves the following nonlinear ordinary differential equation:
$$
    i a' + \Omega a + \bar{a} = 3 \gamma_0 |a|^2 a + 3 \sigma^2 |a|^4 a\,.
$$
Converting the function $a(X)$ to polar coordinates,
\begin{equation}
   a(X) = \sqrt{Q(X)} \exp[-i\Theta(X)/2]\,,
\end{equation}
we obtain the second-order system,
\begin{align}
        Q' &= -2 Q \sin\Theta\,, \notag \\
        \Theta' &= -2 \Omega - 2 \cos\Theta + 6 \gamma_0 Q + 6\sigma^2 Q^2\,. \label{dyn}
\end{align}
This system has the first integral
\begin{equation}
    E = -\Omega Q - Q\cos\Theta + \frac{3}{2} \gamma_0 Q^2 + \sigma^2 Q^3 \,,
\end{equation}
where $E = 0$ from the zero boundary conditions $Q(X) \to 0$ as $|X| \to \infty$.

In the remainder of this paper, we consider the case $\gamma_0 = 0$ 
and $\gamma_1 \neq 0$, which shows the effects of Feshbach resonance 
management on the existence and stability of gap solitons. For the 
case $\gamma_0 \neq 0$ and $\gamma_1 = 0$, exact analytical 
solutions for gap solitons are available \cite{desterkesipe} and the 
stability problem has been analyzed numerically (see 
\cite{chugpel}). When $\gamma_0 = 0$, the second-order system 
(\ref{dyn}) reduces to the first-order differential equation,
\begin{equation}
    \label{theta} \Theta' = 4(\Omega + \cos\Theta)\,.
\end{equation}
The function $Q(X)$ is found from $\Theta(X)$ with the relation
\begin{equation}
    \label{Q} Q^2 = \frac{\Omega + \cos \Theta}{\sigma^2} \geq 0\,.
\end{equation}
Using the technique from Appendix A of \cite{chugpel}, we find the
exact analytical solution of Eq.~(\ref{theta}) and obtain
\begin{equation}
    \label{solution-theta} \cos\Theta = \frac{\cosh^2{(2 \beta x)} -
    \gamma \sinh^2 {(2 \beta x)}}{\cosh^2{(2 \beta x)} + \gamma \sinh^2 {(2
    \beta x)}}\,,
\end{equation}
where
$$
    \gamma = \frac{1+ \Omega}{1-\Omega}\,,
    \qquad \beta = \sqrt{1 - \Omega^2}\,,
$$
and $|\Omega| < 1$.  Substituting (\ref{solution-theta}) into
(\ref{Q}) then gives
\begin{equation}
    \label{solution-Q} Q^2 = \frac{1}{\sigma^2} \frac{1 + \Omega}{
    \cosh^2{(2 \beta x)} + \gamma \sinh^2 {(2 \beta x)}}\,,
\end{equation}
so that
\begin{equation} 
        a(X) = \frac{\sqrt[4]{\gamma (\cosh{(4 \beta X)} -
    \Omega)}}{\sqrt{\sigma} [\cosh(2 \beta X) + i \sqrt{\gamma} \sinh(2
    \beta X)]} \label{gap-solitons} 
\end{equation}
and
\begin{equation}    
        |a|^2 = \frac{\beta}{\sigma \sqrt{\cosh(4\beta X) - \Omega}}\,.
\end{equation}
The limit $\Omega \to -1$ yields the small-amplitude gap soliton 
$|a|^2 \to ({\beta}/{\sigma \sqrt{2}}) \; {\rm sech}(2 \beta X)$, 
which satisfies the focusing quintic NLS equation 
(\ref{quantic-NLS}). At this local bifurcation limit, the power of 
the gap soliton has a threshold,
$$
    P \equiv \int_{-\infty}^{\infty} \left( |A|^2 + |B|^2 \right) dX \to 
    P_0 \equiv \frac{\pi}{\sigma \sqrt{2}}\,,
$$
such that the power $P$ is bounded from below by the limiting value 
$P_0$. The opposite limit $\Omega \to +1$ yields the large-amplitude 
(singular) gap soliton $|a|^2 \to ({\beta}/{\sigma \sqrt{2}}) {\rm 
csch}(2 \beta X)$, which satisfies the corresponding defocusing 
quintic NLS equation (\ref{quantic-NLS}). Thus, in accordance with 
the asymptotic reduction to the quintic NLS equation 
(\ref{quantic-NLS}), the family of gap soliton solutions of the 
coupled-mode system (\ref{cme}) bifurcates from the lower spectral 
band ($\Omega = -1$) and terminates at the upper spectral band 
($\Omega = +1$).

\subsection{Stability} \label{stab}

The spectral stability of the gap soliton (\ref{gap-solitons}) follows
from the linearization
\begin{equation}
    \label{linearization} \left \{
    \begin{array}{c}
        A(X,T) = e^{-i\Omega T} \left( a(X) + U_1(X) e^{\lambda T} \right) \\
        \bar{A}(X,T) = e^{i\Omega T} \left( \bar{a}(X) + U_2(X) e^{\lambda T} \right) \\
        B(X,T) = e^{-i\Omega T} \left( \bar{a}(X)+ U_3(X) e^{\lambda T} \right) \\
        \bar{B}(X,T) = e^{i\Omega T} \left( a(X)+ U_4(X) e^{\lambda T}
        \right)
    \end{array}
    \right.\,.
\end{equation}
The vector ${\bf U} = (U_1,U_2,U_3,U_4)^T$ solves the linear
eigenvalue problem,
\begin{equation}
    \label{linearized-Ham-system} H_{\omega} {\bf U} = i \lambda s {\bf U}\,,
\end{equation}
where $s = \mbox{diag}(1,-1,1,-1)$ is a diagonal matrix.  The
linearized energy operator $H_{\omega}$ has the form
\begin{equation}
    H_{\omega} = D(\partial_X) + \mathcal{V}(X)\,,
\end{equation}
where
\begin{equation*}
    \label{operator-D} D = \left( \begin{array}{cccc} -\Omega - i
    \partial_X & 0 & -1 & 0 \\ 0 & -\Omega + i \partial_X & 0 & -1 \\
    -1 & 0 & -\Omega + i \partial_X & 0 \\
    0 & -1 & 0 & -\Omega - i \partial_X \end{array}\right)
\end{equation*}
and
\begin{equation*}
    \label{operator-V} \mathcal{V} = \sigma^2 \left(
    \begin{array}{cccc} 5 |a|^4 & 2 |a|^2 a^2 &  4 |a|^2 a^2 &  4 |a|^4 \\
        2 |a|^2 \bar{a}^2 & 5 |a|^4 & 4 |a|^4 &  4 |a|^2 \bar{a}^2 \\
        4 |a|^2 \bar{a}^2 & 4 |a|^4 &  5 |a|^4 &  2 |a|^2 \bar{a}^2 \\
        4 |a|^4 & 4 |a|^2 a^2 &  2 |a|^2 a^2 &  5 |a|^4
    \end{array}\right)\,.
\end{equation*}
Using the block-diagonalization method from \cite{chugpel}, we employ
the orthogonal similarity matrix
$$
    S = \frac{1}{\sqrt{2}} \left (
    \begin{array}{cccc}1 & 0 & 1 & 0 \\ 0 & 1 & 0 & 1 \\ 0 & 1 & 0 & -1 \\
    1 & 0 & -1 & 0 \end{array}\right)
$$
that simultaneously block-diagonalizes the energy operator
$H_{\omega}$,
\begin{equation}
    \label{block1}
    S^{-1} H_{\omega} S = \left ( \begin{array}{cc} H_+ & 0 \\
    0 & H_- \end{array}\right) \equiv H,
\end{equation}
and the linearized operator $s H_{\omega}$,
\begin{equation}
    \label{block2} S^{-1} s H_{\omega} S = s \left(
    \begin{array}{cc} 0 & H_- \\ H_+ & 0 \end{array}\right) \equiv i L\,,
\end{equation}
where $H_{\pm}$ are two-by-two Dirac operators:
\begin{align}
    \label{dirac1} H_+ &= \left(
    \begin{array}{cc} -\Omega - i \partial_X + 9 \sigma^2|a|^4 & 6 \sigma^2|a|^2 a^2 - 1\\
        6 \sigma^2|a|^2 \bar{a}^2 - 1 & -\Omega + i \partial_X + 9\sigma^2|a|^4
    \end{array} \right)\,,  \\
    \label{dirac2} H_- &= \left(
    \begin{array}{cc}
        -\Omega - i \partial_X + \sigma^2|a|^4 & 1 - 2 \sigma^2|a|^2 a^2\\
        1 - 2\sigma^2|a|^2 \bar{a}^2 & -\Omega + i \partial_X + \sigma^2|a|^4
    \end{array} \right)\,.
\end{align}
Eigenvalues of the operators $L$, $H_+$, and $H_-$ are detected 
numerically with the Chebyshev interpolation method \cite{chugpel}.  
The main advantage of the Chebyshev grid is that clustering of the 
grid points occurs near the end points of the interval.  This prevents the appearance of spurious complex eigenvalues that may otherwise arise from the discretization of the continuous spectrum.  Moreover, by using the block-diagonalization in (\ref{block1})--(\ref{block2}), we are able 
to reduce the memory constraints and double the speed of the numerical 
computations (see the details in \cite{chugpel}). 

\begin{figure}
 \centerline{
                \includegraphics[width=0.25\textwidth]{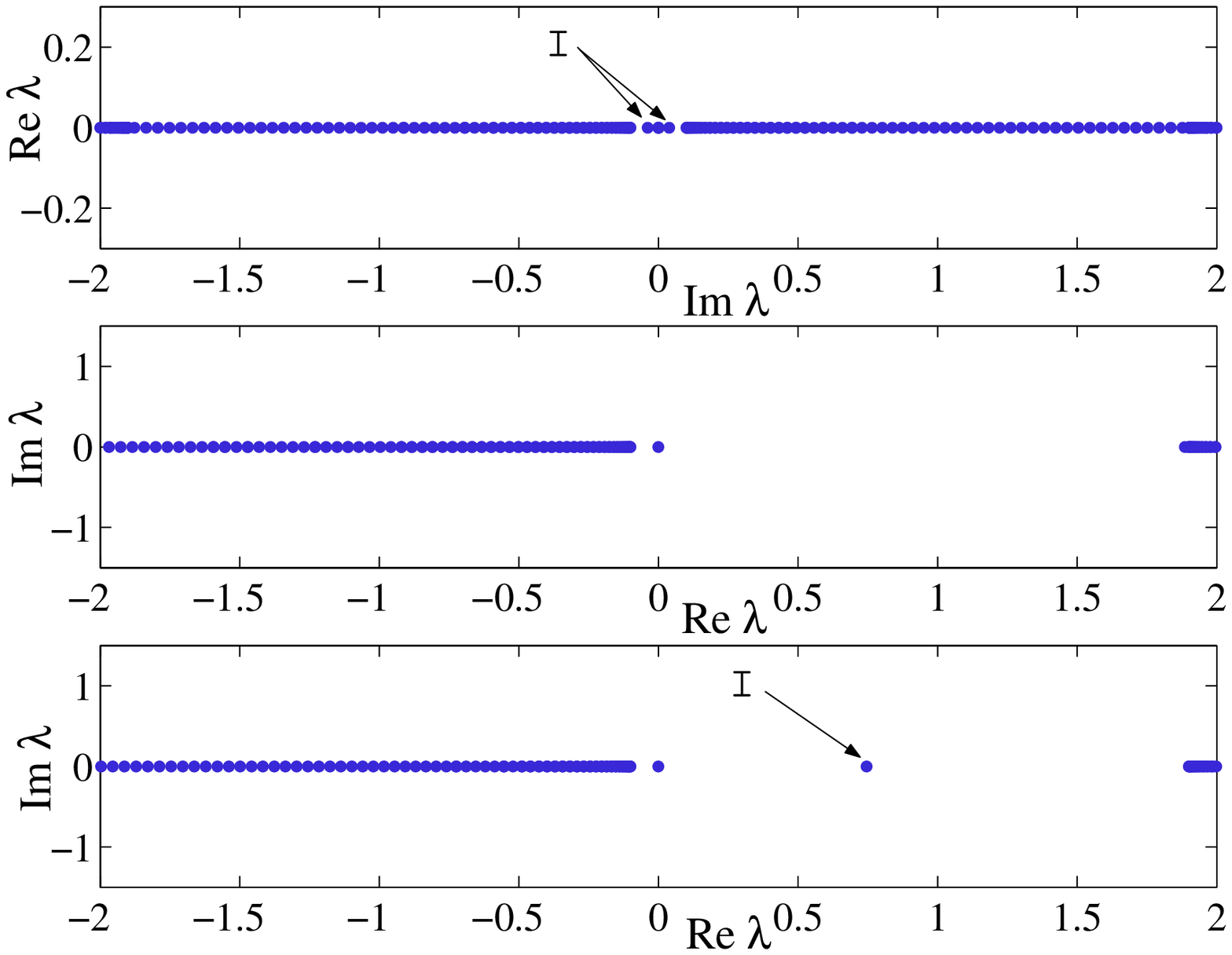}
                \includegraphics[width=0.25\textwidth]{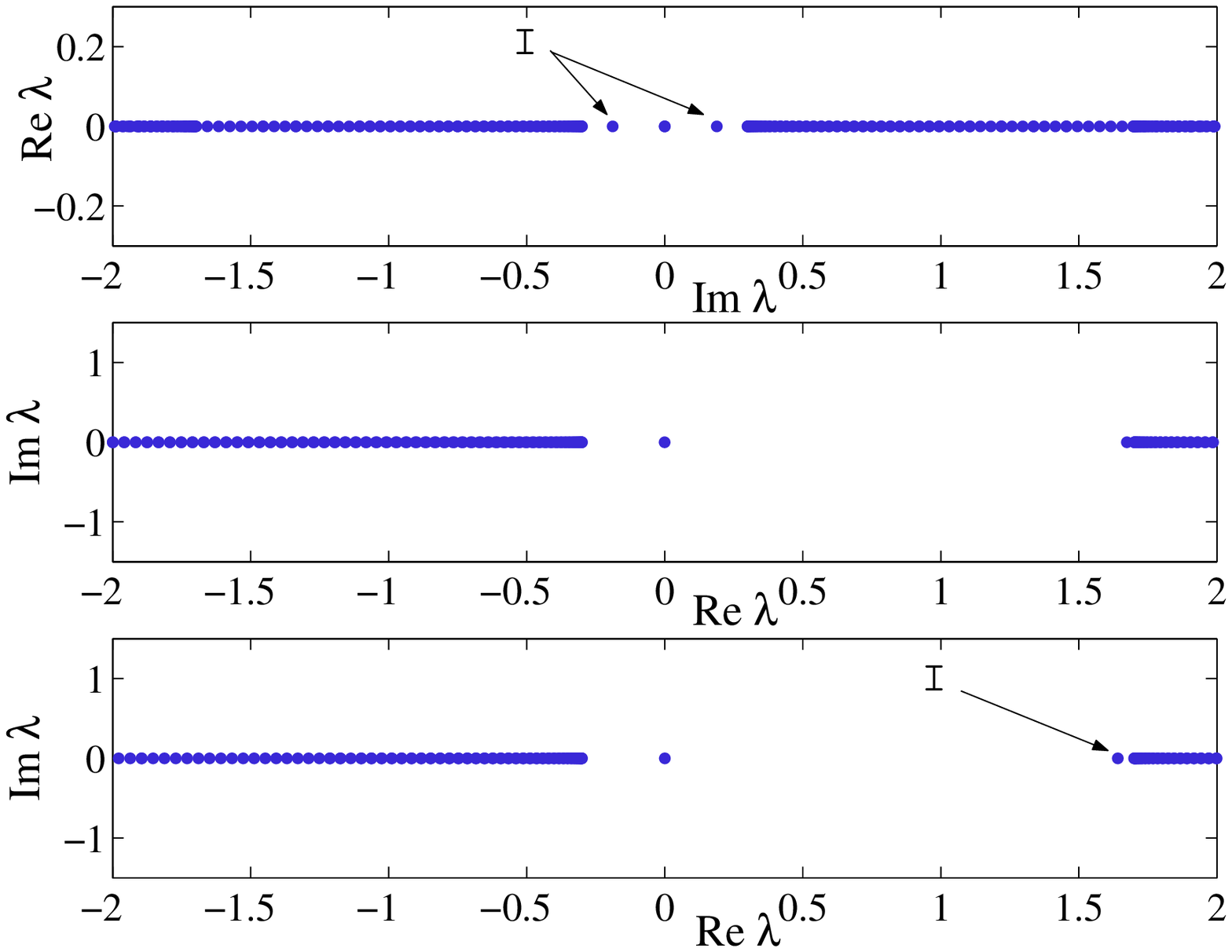}}
 \centerline{
                \includegraphics[width=0.25\textwidth]{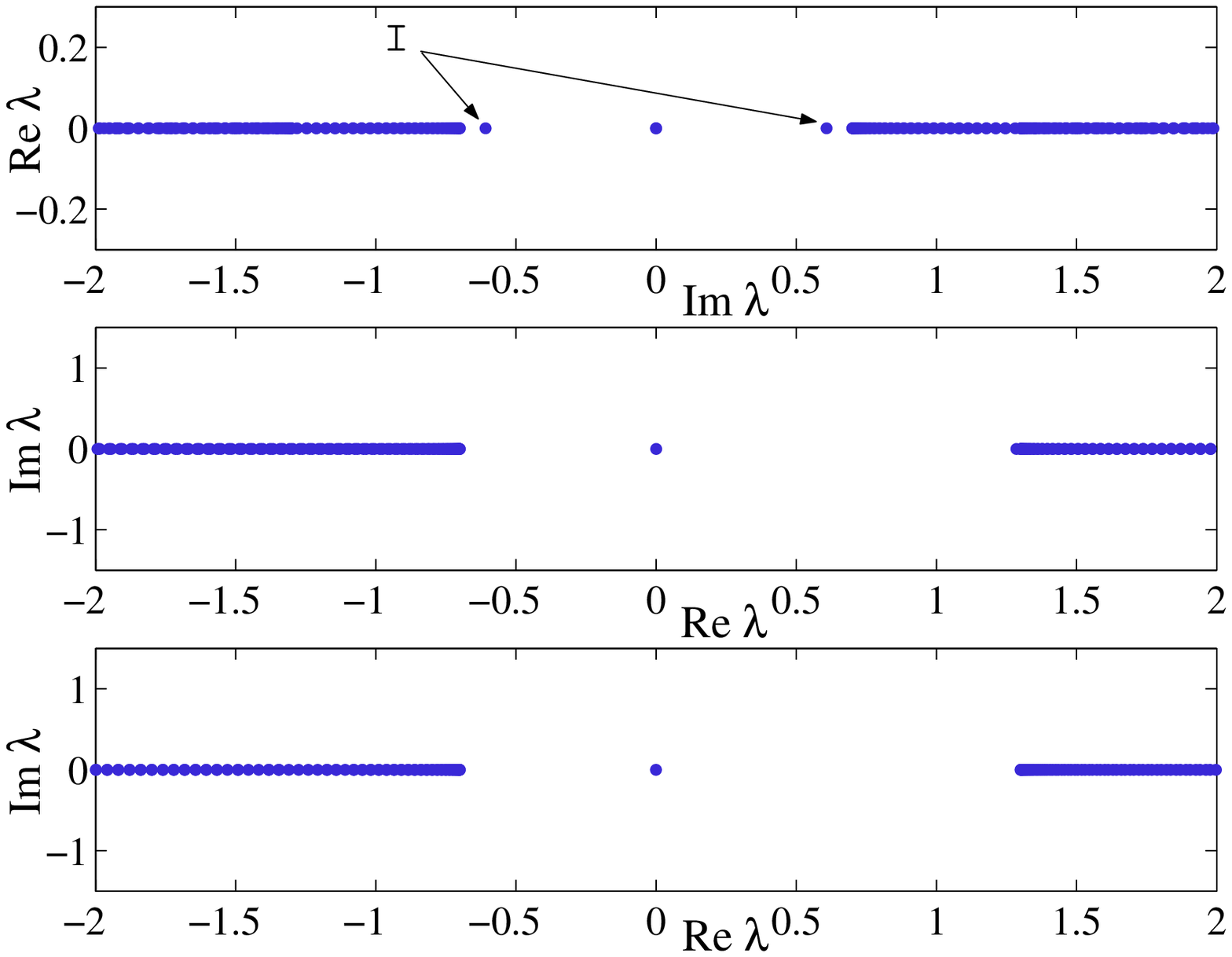}
                \includegraphics[width=0.25\textwidth]{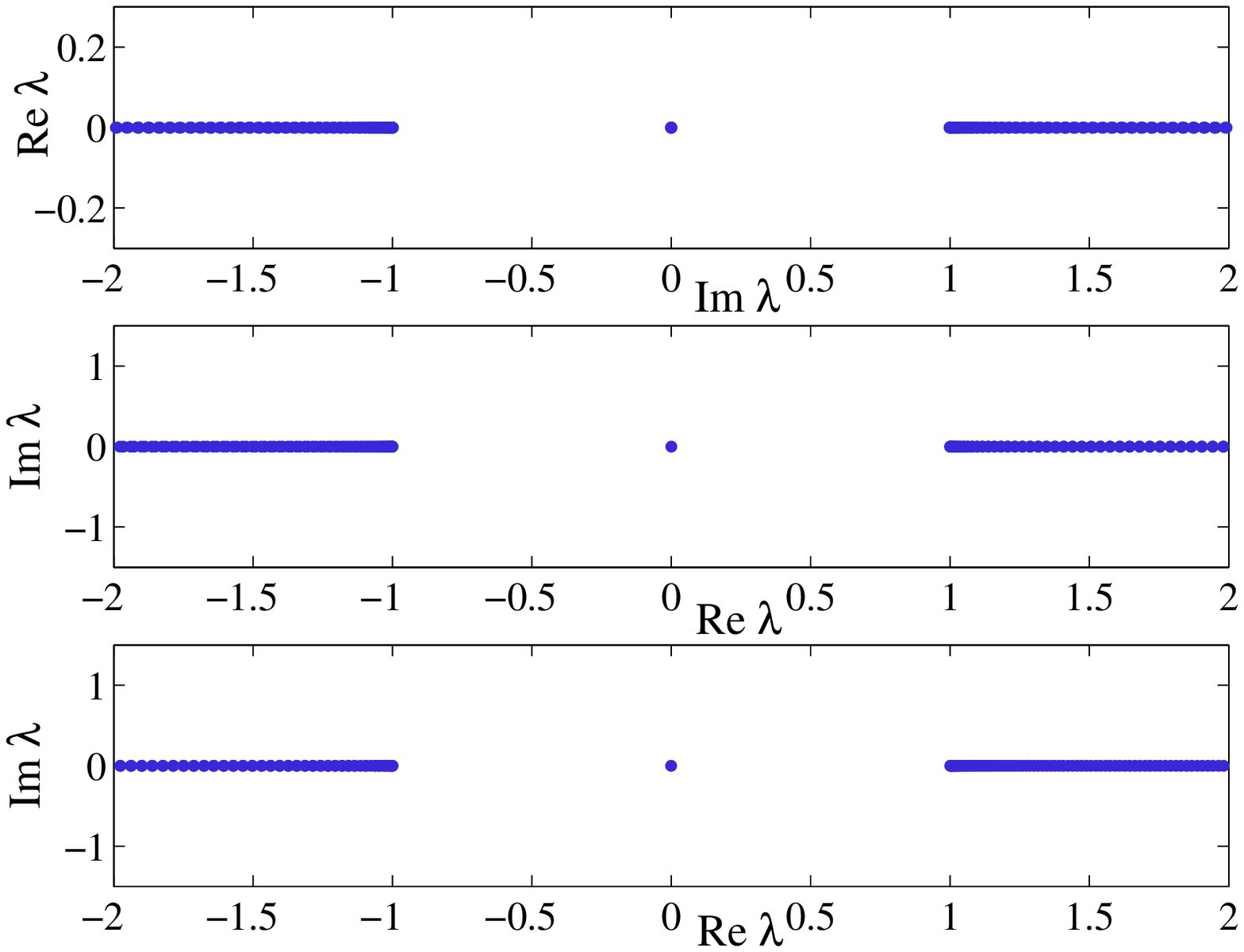}}
 \centerline{
                \includegraphics[width=0.25\textwidth]{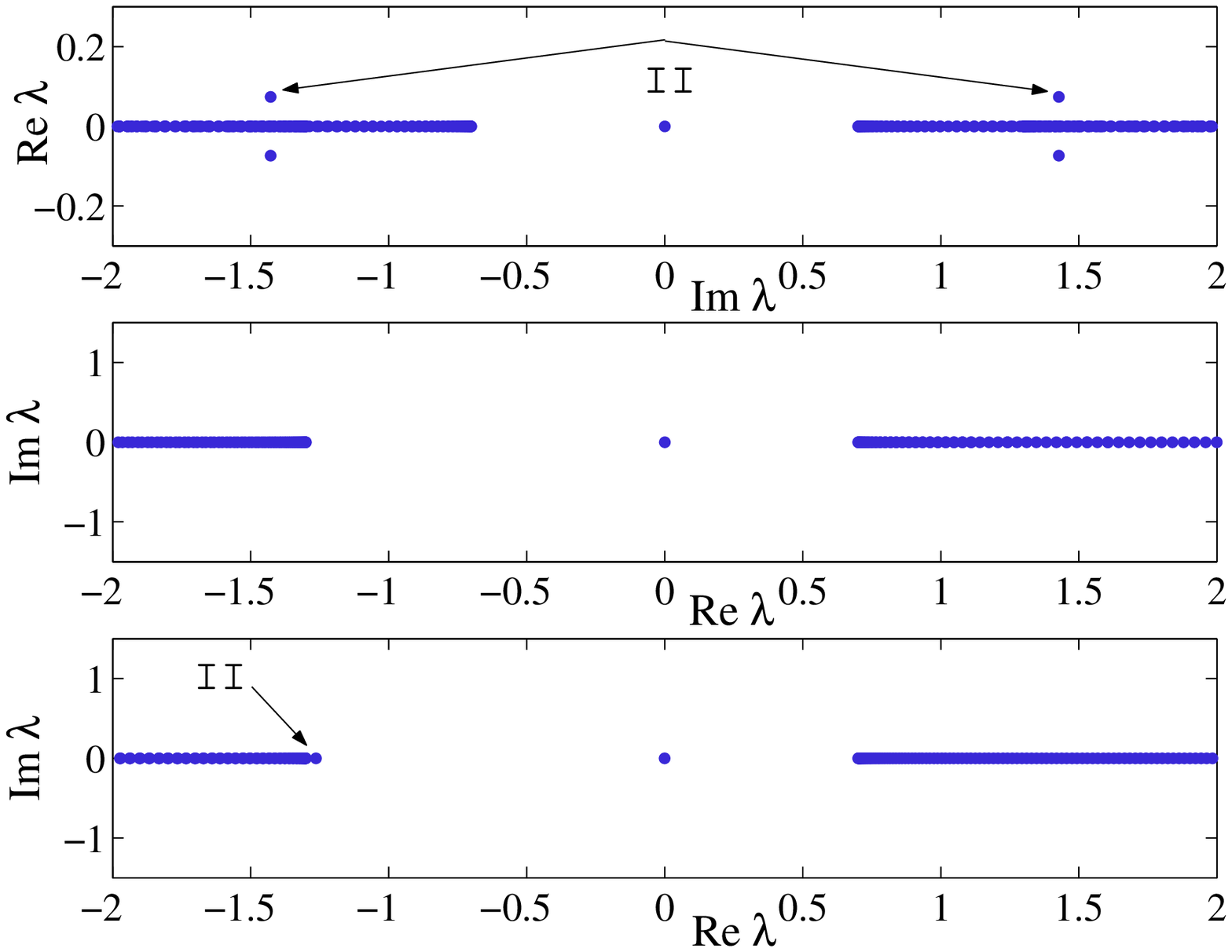}
                \includegraphics[width=0.25\textwidth]{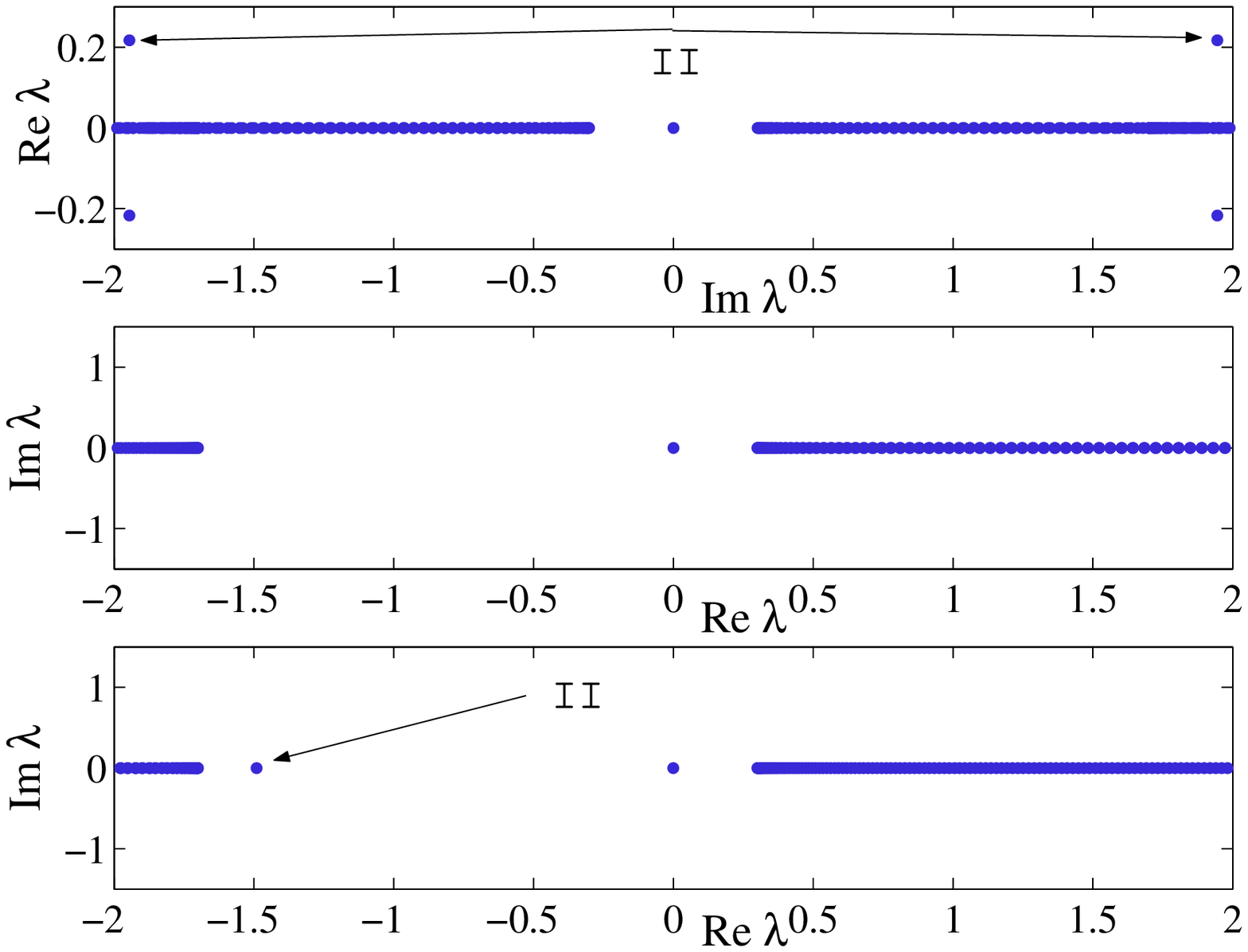}}

\caption{(Color online) Eigenvalues and instability bifurcations for 
the operators $L$, $H_+$, and $H_-$ in (\ref{block2}), (\ref{dirac1}, and 
(\ref{dirac2}).  The parameter values are  $\Omega = -0.9$ (top left), $\Omega = 
-0.7$ (top right), $\Omega = -0.3$ (middle left), $\Omega = 0.0$ (middle right), $\Omega = 0.3$ (bottom left), and $\Omega = 0.7$ (bottom right).}\label{eval}
\end{figure}

The numerical eigenvalues of the operators $L$, $H_+$, and $H_-$ are
displayed in Fig.~\ref{eval} for six different values of the 
parameter $\Omega$. When $\Omega$ is close to the local bifurcation 
threshold (e.g., for $\Omega = -0.9$), the operator $L$ has a 
four-dimensional kernel at $\lambda = 0$ and a pair of small purely 
imaginary eigenvalues near $\lambda = 0$. The pair of purely 
imaginary eigenvalues originates from the six-dimensional kernel of 
the linearized quintic NLS equation (\ref{quantic-NLS}) (see, e.g., 
Ref.~\cite{pelkivafan}). In this case, the operator $H_+$ has no 
isolated non-zero eigenvalues, whereas the operator $H_-$ has a 
simple isolated non-zero eigenvalue. The eigenvalue of $H_-$ still 
exists at $\Omega = -0.7$, but it disappears before $\Omega = -0.3$ 
because it collides with the end point of the continuous spectrum of 
$H_-$. The pair of eigenvalues of $L$ survives at $\Omega = -0.3$, 
but it disappears at $\Omega = 0$ because it collides with the end 
points of the continuous spectrum of $L$. When $\Omega > 0$, the 
pair of complex eigenvalues bifurcates in the spectrum of $L$ from 
the end points of the continuous spectrum of $L$. The complex 
eigenvalues of $L$ bifurcate simultaneously with a simple isolated 
non-zero eigenvalue of the operator $H_-$. The pair of complex 
eigenvalues of $L$ and the isolated non-zero eigenvalue of $H_-$ 
persist for larger values of $\Omega$ (e.g., for $\Omega = 0.3$ and 
$\Omega = 0.7$).  The eigenvalues just discussed are labeled in the 
figure as ${\rm I}$ and ${\rm II}$.

In sum, the gap solitons of the coupled-mode system 
(\ref{cme}) with $\gamma_0 = 0$ are spectrally stable for $\Omega < 0$ 
and spectrally unstable due to complex unstable eigenvalues for 
$\Omega > 0$. This behavior is similar to the stability analysis of 
the coupled-mode system (\ref{cme}) with $\gamma_0 \neq 0$ and $\gamma_1 = 0$ 
(see \cite{chugpel} and references therein).

\section{Numerical simulations} \label{numerics}

To confirm the accuracy of the coupled-mode theory, we corroborate the instability of gap solitons for $\Omega > 0$ predicted from the analysis of the system (\ref{cme}) with full numerical simulations of the averaged GP equation (\ref{avg2}). We consider the trigonometric management function $\gamma(\tau) = \cos(2\pi \tau)$ and fix the other parameters as follows: $\gamma_0 = 0$, $\gamma_1 = 1/\sqrt{2}$, $V_0 = - 1$, $\omega = 1$, and $\epsilon = 0.1$. We vary the parameter $\Omega$ for the gap soliton solutions, focusing on computations with $\Omega = -0.5$ and $\Omega = 0.5$. According to analysis of the coupled-mode equations (\ref{cme}), the value $\Omega = -0.5$ corresponds to the case of stable gap solitons, and the value $\Omega = 0.5$ corresponds to the case of unstable gap solitons. The initial condition for all simulations is selected to be a perturbation of the leading-order two-wave approximation (\ref{two-wave}) with the gap soliton solutions (\ref{gap-solitons}).

We integrated the averaged GP equation (\ref{avg2}) using a finite-difference approximation (with 480 grid points) in space and a Runge-Kutta integration scheme (with step size $h = 0.001$) in time.  The perturbations of the initial gap solitons are of the same functional form as the gap solitons but with larger amplitudes.

\begin{figure}[p]
                \centerline{
                \includegraphics[width=0.25\textwidth]{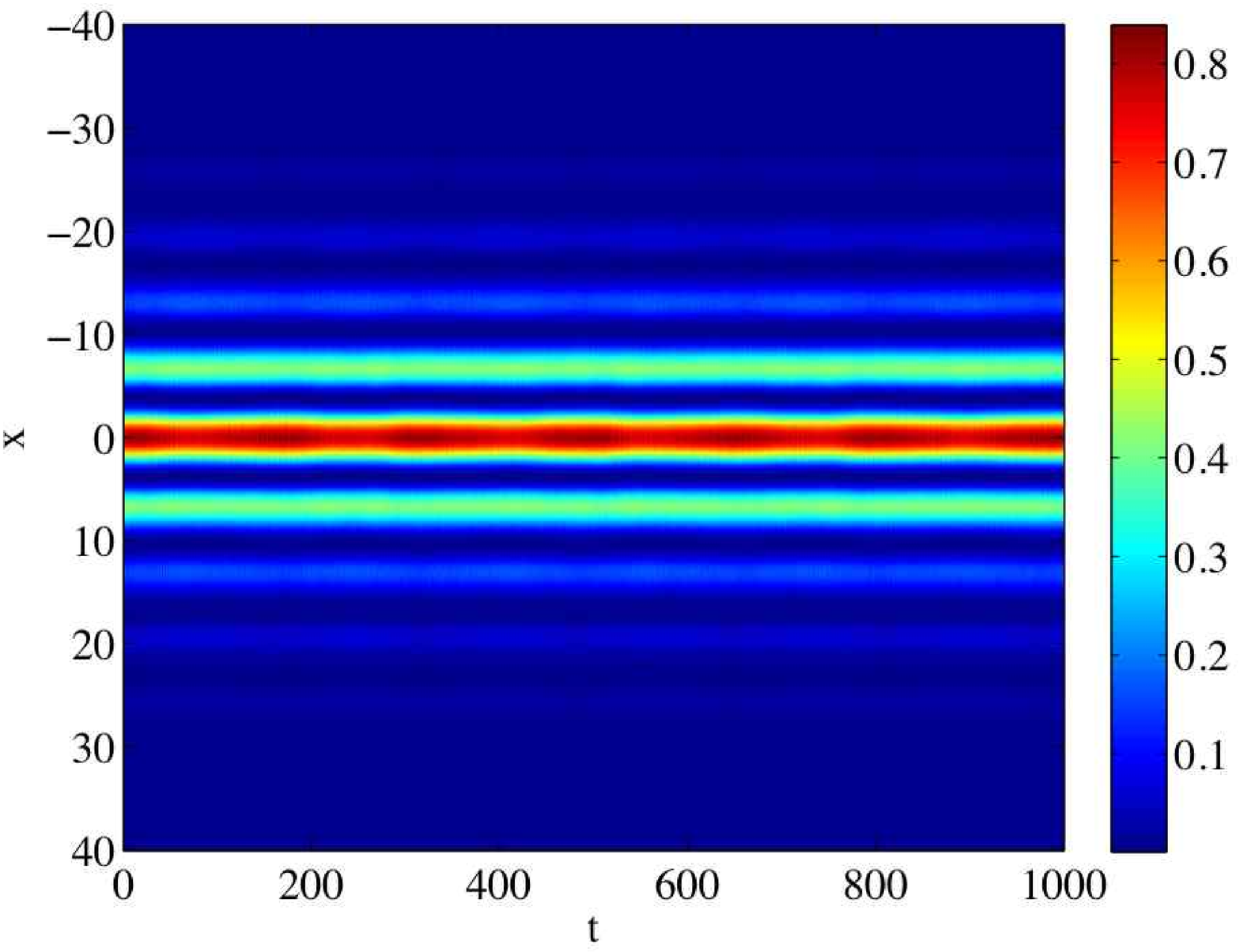}
                \includegraphics[width=0.25\textwidth]{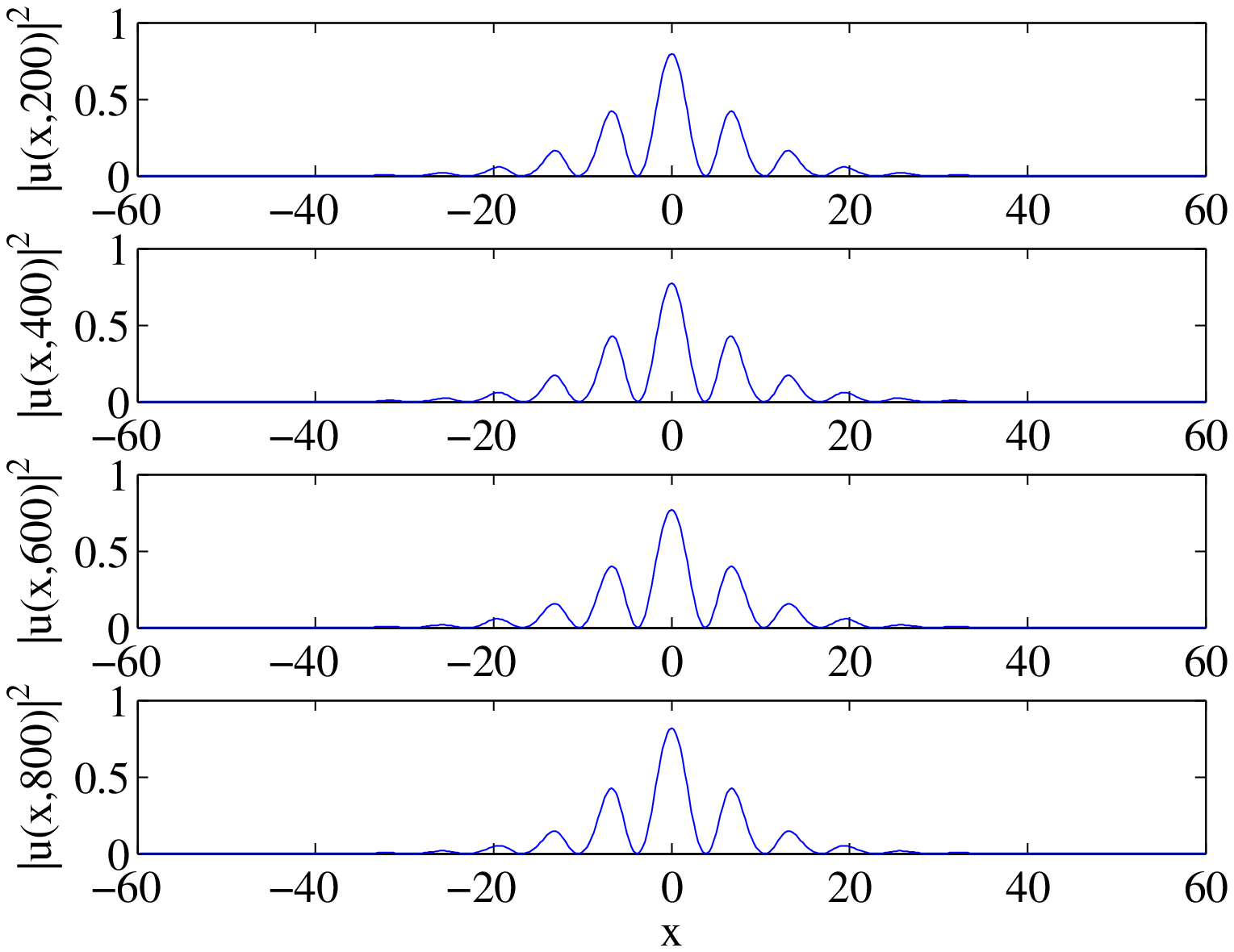}}
                \caption{(Color online) Stable time-evolution of a gap soliton with $\Omega = -0.5$ in the averaged GP equation (\ref{avg2}).  (Left) Spatio-temporal evolution of $|u(x,t)|^2$.  (Right) Spatial profiles of $|u(x,t)|^2$ for $t = 0$, $t = 200$, $t = 400$, and $t = 800$.} \label{minus}
\end{figure}

Figure \ref{minus} shows the time-evolution of a stable gap soliton with $\Omega = -0.5$. The stationary gap soliton persists in the full dynamics of the averaged GP equation (\ref{avg2}), in agreement with the stability analysis of the coupled-mode system (\ref{cme}).

\begin{figure}
                \centerline{
                \includegraphics[width=0.25\textwidth]{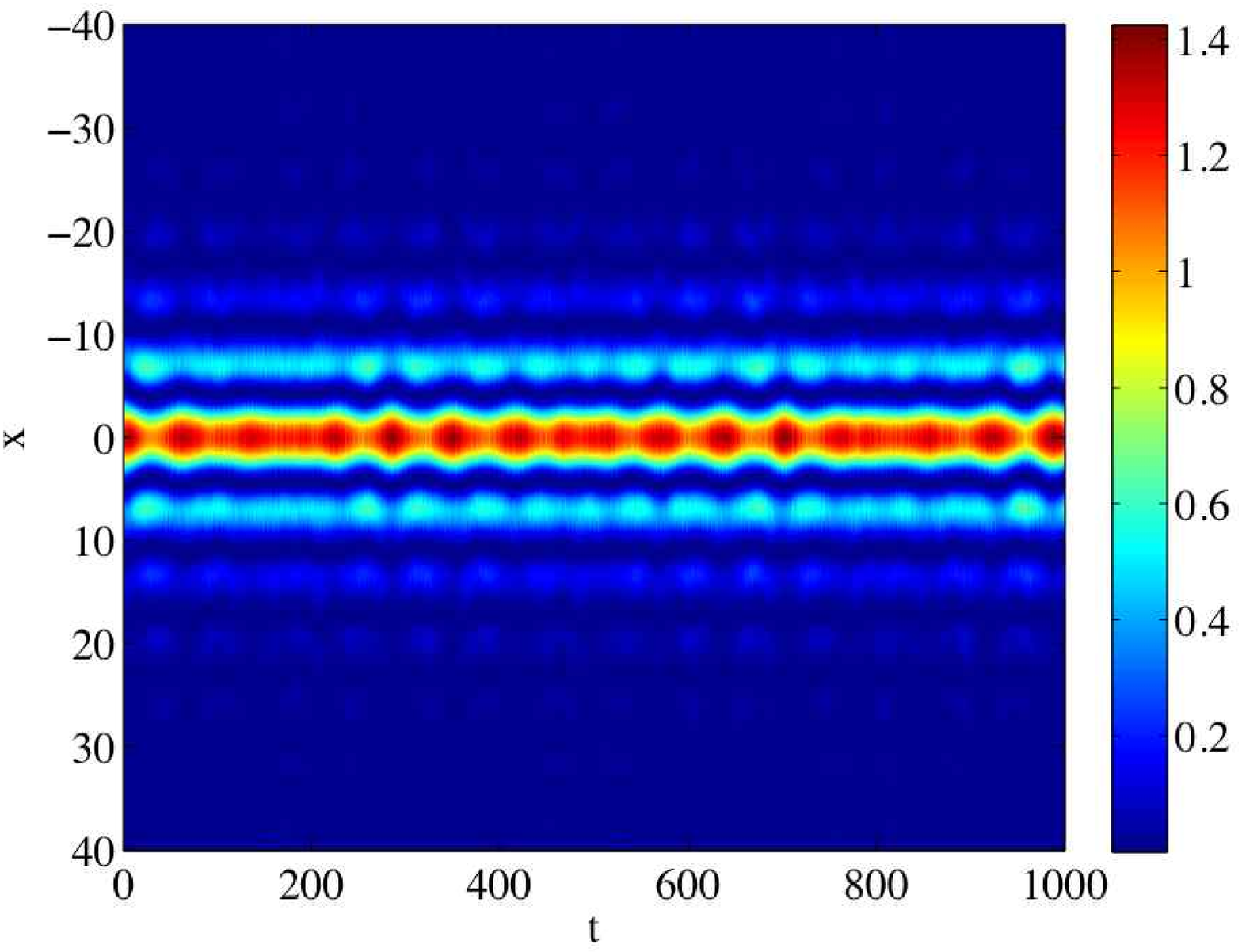}
                \includegraphics[width=0.25\textwidth]{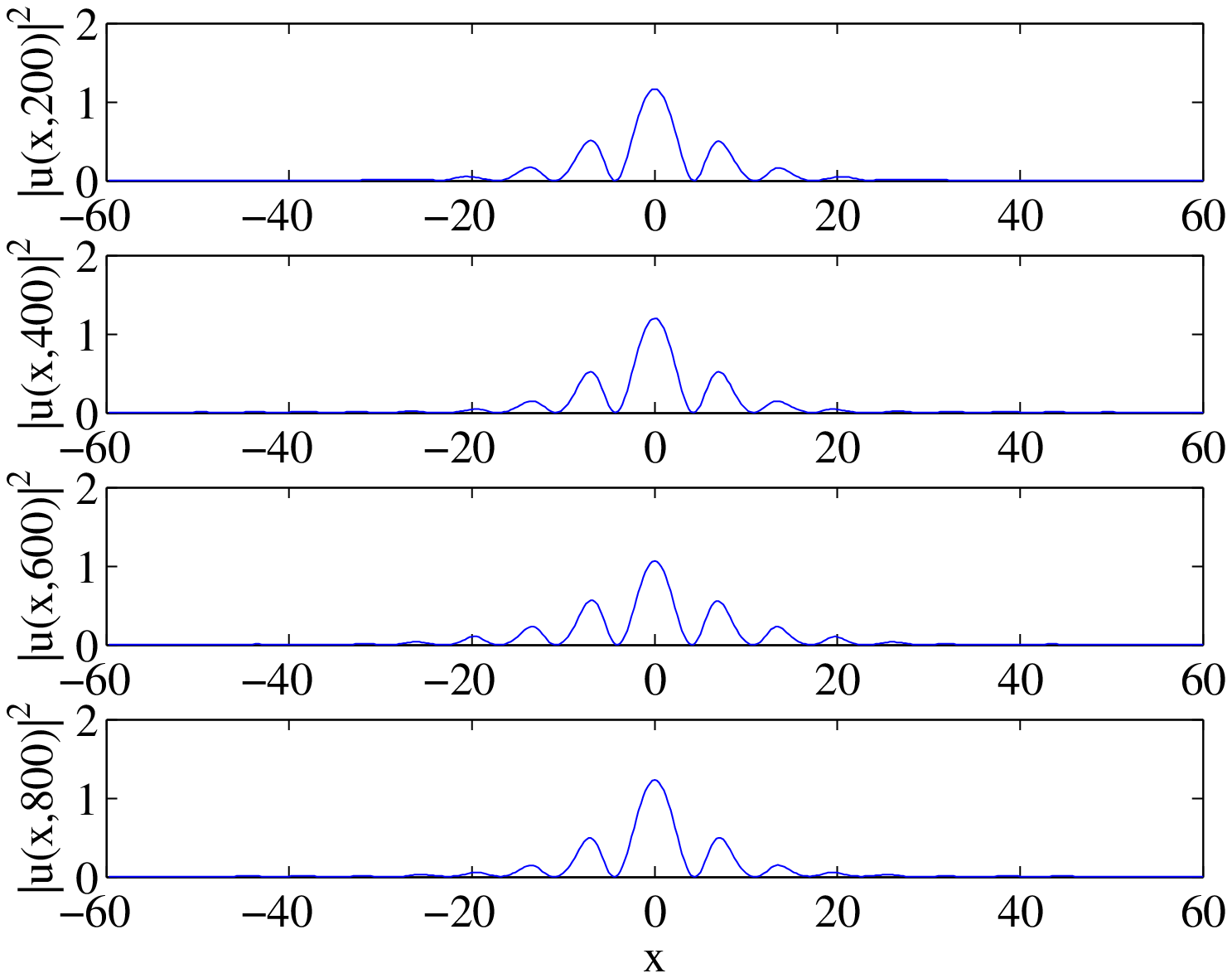}}
                \caption{(Color online) Unstable time-evolution of a gap soliton with $\Omega = 0.5$ in the averaged GP equation (\ref{avg2}).  (Left) Spatio-temporal evolution of $|u(x,t)|^2$.  (Right) Spatial profiles of $|u(x,t)|^2$ for $t = 0$, $t = 200$, $t = 400$, and $t = 800$.} \label{plus}
\end{figure}

Figure \ref{plus} illustrates the dynamics of an unstable gap soliton with $\Omega = 0.5$.  We observe an asymmetric beating between different localized waveforms.  The localized wave is not destroyed in the unstable case, but rather undergoes shape distortions due to the oscillatory instability. This behavior agrees with the stability analysis in the coupled-mode system (\ref{cme}), as the unstable eigenvalues of gap solitons with $\Omega > 0$ are complex-valued (see Fig.~\ref{eval}).   While the perturbation used in Fig.~\ref{plus} is  large, note that a perturbation with the same percentage difference in the soliton amplitude does not lead to an instability in Fig.~\ref{minus}.

\begin{figure}
                \centerline{
                \includegraphics[width=0.25\textwidth]{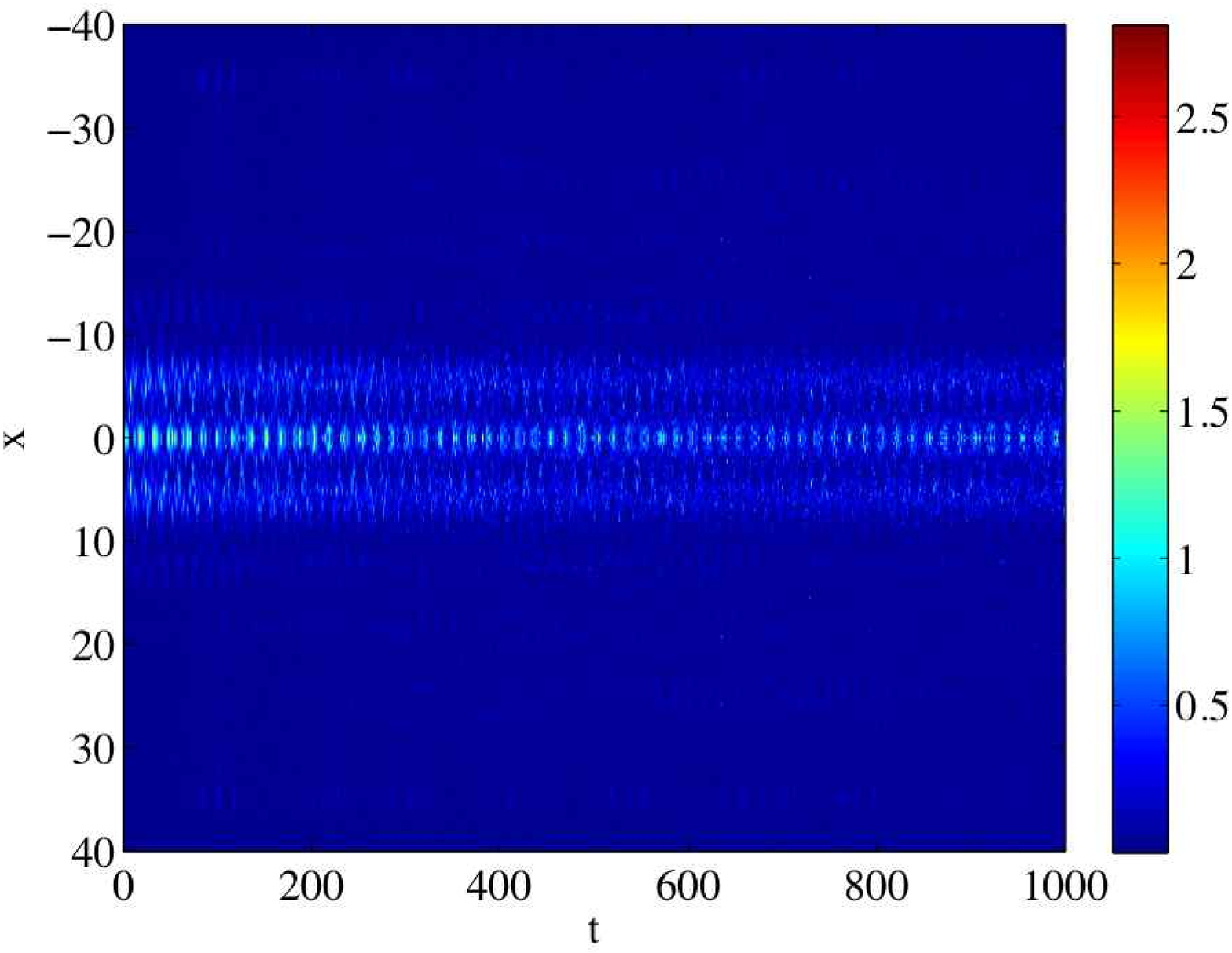}
                \includegraphics[width=0.25\textwidth]{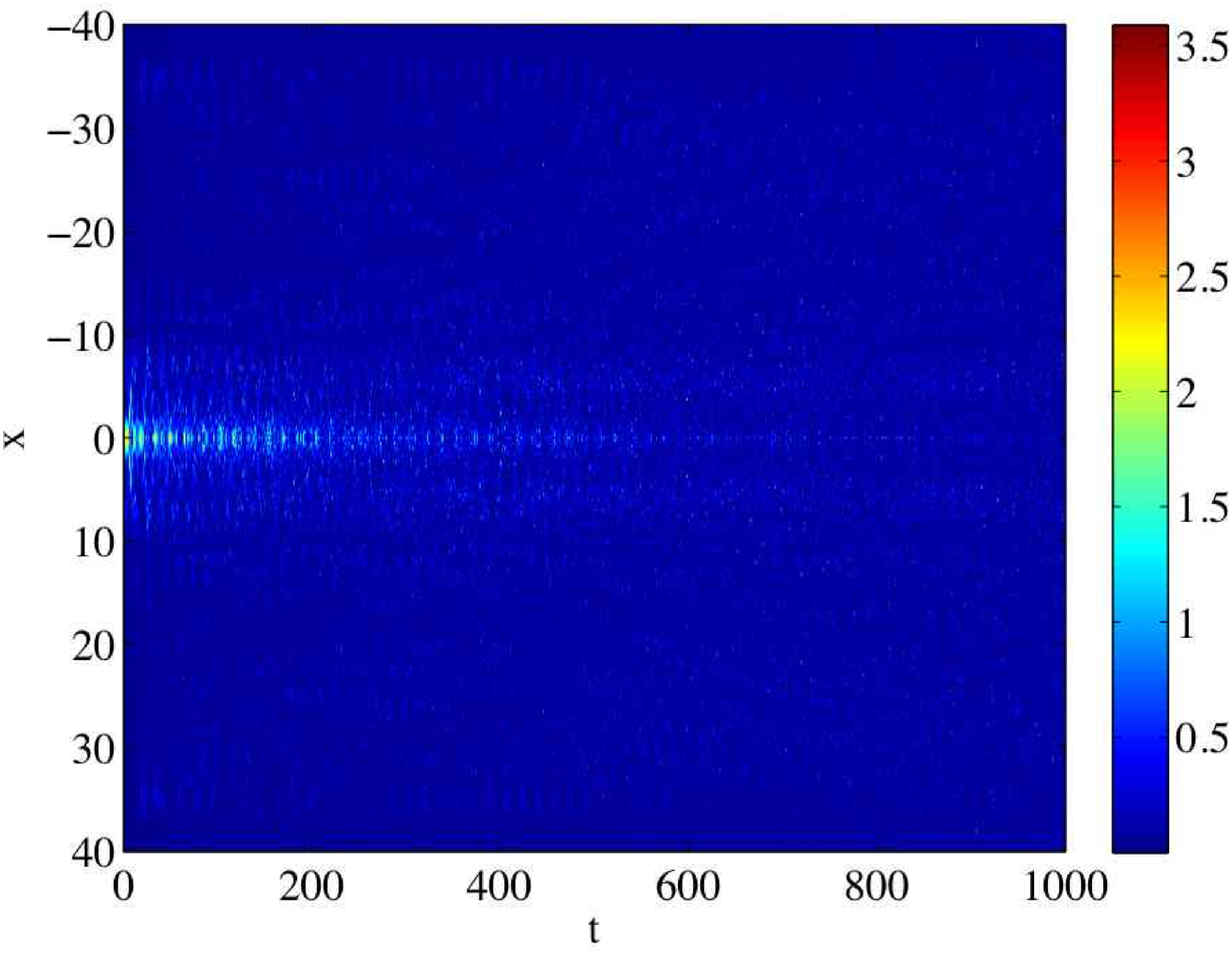}}
                \caption{(Color online) Spatio-temporal evolution of $|\psi(x,t)|^2$, with initial conditions given by the constructed gap solitons, under the dynamics of the full GP equation (\ref{GP}).  (Left) $\Omega = -0.5$.  (Right) $\Omega = 0.5$.} \label{original}
\end{figure}

We also examined the dynamics of the solutions constructed using coupled-mode theory under full numerical simulations of the original GP equation (\ref{GP}).  In these simulations, which were also performed using a finite-difference approximation (with 480 grid points) in space and a Runge-Kutta integration scheme (with step size $h = 0.001$) in time, the external potential includes contributions from a small harmonic trap in addition to the optical lattice in (\ref{four}).  Hence, the potential in these simulations is given by $\hat{V}(x) = V(x) + V_h x^2$.  The initial conditions and parameters are the same as for the simulations of the averaged GP equation, with the additional values $\varepsilon = 0.25$ and $V_h = 0.01$.  In Fig.~\ref{original}, we show the spatio-temporal evolution of the constructed gap solitons under equation (\ref{GP}).  As expected, the solitons that we found to be stable persist longer with respect to the full GP dynamics than those determined to be unstable.

\section{Conclusions}

In conclusion, we studied Feshbach resonance management for gap 
solitons in Bose-Einstein condensates trapped in optical lattice 
potentials.  We applied an envelope wave approximation to the 
averaged Gross-Pitaevsky equation with a periodic potential to yield 
coupled-mode equations describing the slow BEC dynamics in the first 
spectral gap of the optical lattice.  We derived exact analytical 
expressions for gap solitons with zero mean scattering length.  In 
this situation, Feshbach resonances are employed to tune a 
condensate between the repulsive and attractive regimes 
(corresponding to their usual experimental application).  Applying 
Chebyshev interpolation to the coupled-mode equations, we showed 
that these gap solitons are unstable above the center of the first 
spectral gap (far from the local bifurcation threshold).  We then 
showed with numerical simulations of the averaged Gross-Pitaevsky (GP)
equation that unstable gap solitons exhibit beating between 
different localized shapes, thereby confirming the stability results predicted from the coupled-mode theory.  We corroborated this further with numerical simulations of the original GP equation, which show that the stable gap solitons persist much longer than the unstable ones.

We note that gap solitons in the GP equation with a 
periodic potential (\ref{avg2}) and the coupled-mode system 
(\ref{cme}) in the case of no Feshbach resonance management have 
been studied recently in \cite{pelsukhkiv} and \cite{chugpel}, 
respectively. We can see from comparing the previous and new results 
that Feshbach resonance management leads to a new effect with 
respect to the existence of the power threshold near the local 
bifurcation limit \cite{malfesh05}. On the other hand, 
there are not many differences in the stability results, as Feshbach 
resonance management does not stabilize gap solitons far from the 
threshold of local bifurcations (which are known to be unstable 
\cite{chugpel}).  We have also confirmed that Feshbach resonance 
management leads to defocusing effects on the propagation of gap 
solitons, which is relevant for blow-up arrest in multi-dimensional 
problems \cite{kevstefpel}.

\section{Acknowledgements}

We gratefully acknowledge Jit Kee Chin, Randy Hulet, Panos Kevrekidis, Boris Malomed, and Steve Rolston for useful discussions about this project. The code for numerical simulations of the NLS was modified from the code of Panos Kevrekidis.  M.A.P. acknowledges support from the NSF VIGRE program and the Gordon and Betty Moore Foundation.  M. Ch. was supported by the ShacrNet and NSERC graduate scholarships. D. P. was supported by the NSERC Discovery and PREA grants.


\clearpage

\begin{thebibliography}{47}
\expandafter\ifx\csname natexlab\endcsname\relax\def\natexlab#1{#1}\fi
\expandafter\ifx\csname bibnamefont\endcsname\relax
  \def\bibnamefont#1{#1}\fi
\expandafter\ifx\csname bibfnamefont\endcsname\relax
  \def\bibfnamefont#1{#1}\fi
\expandafter\ifx\csname citenamefont\endcsname\relax
  \def\citenamefont#1{#1}\fi
\expandafter\ifx\csname url\endcsname\relax
  \def\url#1{\texttt{#1}}\fi
\expandafter\ifx\csname urlprefix\endcsname\relax\def\urlprefix{URL }\fi
\providecommand{\bibinfo}[2]{#2}
\providecommand{\eprint}[2][]{\url{#2}}


\bibitem[{\citenamefont{Dalfovo et~al.}(1999)\citenamefont{Dalfovo, Giorgini,
  Pitaevskii, and Stringari}}]{stringari}
\bibinfo{author}{\bibfnamefont{F.}~\bibnamefont{Dalfovo}},
  \bibinfo{author}{\bibfnamefont{S.}~\bibnamefont{Giorgini}},
  \bibinfo{author}{\bibfnamefont{L.~P.} \bibnamefont{Pitaevskii}},
  \bibnamefont{and}
  \bibinfo{author}{\bibfnamefont{S.}~\bibnamefont{Stringari}},
  \bibinfo{journal}{Reviews of Modern Physics} \textbf{\bibinfo{volume}{71}},
  \bibinfo{pages}{463} (\bibinfo{year}{1999}).

\bibitem[{\citenamefont{K\"ohler}(2002)}]{kohler}
\bibinfo{author}{\bibfnamefont{T.}~\bibnamefont{K\"ohler}},
  \bibinfo{journal}{Physical Review Letters} \textbf{\bibinfo{volume}{89}},
  \bibinfo{pages}{210404} (\bibinfo{year}{2002}).
  
\bibitem[{\citenamefont{Anderson and Kasevich}(1998)}]{anderson}
\bibinfo{author}{\bibfnamefont{B.~P.} \bibnamefont{Anderson}} \bibnamefont{and}
  \bibinfo{author}{\bibfnamefont{M.~A.} \bibnamefont{Kasevich}},
  \bibinfo{journal}{Science} \textbf{\bibinfo{volume}{282}},
  \bibinfo{pages}{1686} (\bibinfo{year}{1998}).
  
\bibitem[{\citenamefont{Orzel et~al.}(2001)\citenamefont{Orzel, Tuchman,
  Fenselau, Yasuda, and Kasevich}}]{squeeze}
\bibinfo{author}{\bibfnamefont{C.}~\bibnamefont{Orzel}},
  \bibinfo{author}{\bibfnamefont{A.~K.} \bibnamefont{Tuchman}},
  \bibinfo{author}{\bibfnamefont{M.~L.} \bibnamefont{Fenselau}},
  \bibinfo{author}{\bibfnamefont{M.}~\bibnamefont{Yasuda}}, \bibnamefont{and}
  \bibinfo{author}{\bibfnamefont{M.~A.} \bibnamefont{Kasevich}},
  \bibinfo{journal}{Science} \textbf{\bibinfo{volume}{291}},
  \bibinfo{pages}{2386} (\bibinfo{year}{2001}).

\bibitem[{\citenamefont{Morsch et~al.}(2001)\citenamefont{Morsch, M\"uller,
  Cristiani, Ciampini, and Arimondo}}]{morsch}
\bibinfo{author}{\bibfnamefont{O.}~\bibnamefont{Morsch}},
  \bibinfo{author}{\bibfnamefont{J.~H.} \bibnamefont{M\"uller}},
  \bibinfo{author}{\bibfnamefont{M.}~\bibnamefont{Cristiani}},
  \bibinfo{author}{\bibfnamefont{D.}~\bibnamefont{Ciampini}}, \bibnamefont{and}
  \bibinfo{author}{\bibfnamefont{E.}~\bibnamefont{Arimondo}},
  \bibinfo{journal}{Physical Review Letters} \textbf{\bibinfo{volume}{87}},
  \bibinfo{pages}{140402} (\bibinfo{year}{2001}).

\bibitem[{\citenamefont{Machholm et~al.}(2004)\citenamefont{Machholm, Nicolin,
  Pethick, and Smith}}]{pethick2}
\bibinfo{author}{\bibfnamefont{M.}~\bibnamefont{Machholm}},
  \bibinfo{author}{\bibfnamefont{A.}~\bibnamefont{Nicolin}},
  \bibinfo{author}{\bibfnamefont{C.~J.} \bibnamefont{Pethick}},
  \bibnamefont{and} \bibinfo{author}{\bibfnamefont{H.}~\bibnamefont{Smith}},
  \bibinfo{journal}{Physical Review A} \textbf{\bibinfo{volume}{69}},
  \bibinfo{pages}{043604} (\bibinfo{year}{2004}).

\bibitem[{\citenamefont{Porter and Cvitanovi\'c}(2004)}]{mapbecprl}
\bibinfo{author}{\bibfnamefont{M.~A.} \bibnamefont{Porter}} \bibnamefont{and}
  \bibinfo{author}{\bibfnamefont{P.}~\bibnamefont{Cvitanovi\'c}},
  \bibinfo{journal}{Physical Review E} \textbf{\bibinfo{volume}{69}},
  \bibinfo{pages}{047201} (\bibinfo{year}{2004}).

\bibitem[{\citenamefont{Smerzi et~al.}(2002)\citenamefont{Smerzi, Trombettoni,
  Kevrekidis, and Bishop}}]{smerzi}
\bibinfo{author}{\bibfnamefont{A.}~\bibnamefont{Smerzi}},
  \bibinfo{author}{\bibfnamefont{A.}~\bibnamefont{Trombettoni}},
  \bibinfo{author}{\bibfnamefont{P.~G.} \bibnamefont{Kevrekidis}},
  \bibnamefont{and} \bibinfo{author}{\bibfnamefont{A.~R.}
  \bibnamefont{Bishop}}, \bibinfo{journal}{Physical Review Letters}
  \textbf{\bibinfo{volume}{89}}, \bibinfo{pages}{170402}
  (\bibinfo{year}{2002}).

\bibitem[{\citenamefont{Greiner et~al.}(2002)\citenamefont{Greiner, Mandel,
  Esslinger, H\"ansch, and Bloch}}]{mott}
\bibinfo{author}{\bibfnamefont{M.}~\bibnamefont{Greiner}},
  \bibinfo{author}{\bibfnamefont{O.}~\bibnamefont{Mandel}},
  \bibinfo{author}{\bibfnamefont{T.}~\bibnamefont{Esslinger}},
  \bibinfo{author}{\bibfnamefont{T.}~\bibnamefont{H\"ansch}}, \bibnamefont{and}
  \bibinfo{author}{\bibfnamefont{I.}~\bibnamefont{Bloch}},
  \bibinfo{journal}{Nature} \textbf{\bibinfo{volume}{415}}, \bibinfo{pages}{39}
  (\bibinfo{year}{2002}).

\bibitem[{\citenamefont{Vollbrecht et~al.}(2004)\citenamefont{Vollbrecht,
  Solano, and Cirac}}]{voll}
\bibinfo{author}{\bibfnamefont{K.~G.~H.} \bibnamefont{Vollbrecht}},
  \bibinfo{author}{\bibfnamefont{E.}~\bibnamefont{Solano}}, \bibnamefont{and}
  \bibinfo{author}{\bibfnamefont{J.~I.} \bibnamefont{Cirac}},
  \bibinfo{journal}{Physical Review Letters} \textbf{\bibinfo{volume}{93}},
  \bibinfo{pages}{220502} (\bibinfo{year}{2004}).

\bibitem[{\citenamefont{Donley et~al.}(2001)\citenamefont{Donley, Claussen,
  Cornish, Roberts, Cornell, and Weiman}}]{fesh}
\bibinfo{author}{\bibfnamefont{E.~A.} \bibnamefont{Donley}},
  \bibinfo{author}{\bibfnamefont{N.~R.} \bibnamefont{Claussen}},
  \bibinfo{author}{\bibfnamefont{S.~L.} \bibnamefont{Cornish}},
  \bibinfo{author}{\bibfnamefont{J.~L.} \bibnamefont{Roberts}},
  \bibinfo{author}{\bibfnamefont{E.~A.} \bibnamefont{Cornell}},
  \bibnamefont{and} \bibinfo{author}{\bibfnamefont{C.~E.}
  \bibnamefont{Weiman}}, \bibinfo{journal}{Nature}
  \textbf{\bibinfo{volume}{412}}, \bibinfo{pages}{295} (\bibinfo{year}{2001}).

\bibitem[{\citenamefont{Inouye et~al.}(1998)\citenamefont{Inouye, Andrews,
  Stenger, Miesner, Stamper-Kurn, and Ketterle}}]{inouye}
\bibinfo{author}{\bibfnamefont{S.}~\bibnamefont{Inouye}},
  \bibinfo{author}{\bibfnamefont{M.~R.} \bibnamefont{Andrews}},
  \bibinfo{author}{\bibfnamefont{J.}~\bibnamefont{Stenger}},
  \bibinfo{author}{\bibfnamefont{H.~J.} \bibnamefont{Miesner}},
  \bibinfo{author}{\bibfnamefont{D.~M.} \bibnamefont{Stamper-Kurn}},
  \bibnamefont{and} \bibinfo{author}{\bibfnamefont{W.}~\bibnamefont{Ketterle}},
  \bibinfo{journal}{Nature} \textbf{\bibinfo{volume}{392}},
  \bibinfo{pages}{151} (\bibinfo{year}{1998}).

\bibitem[{\citenamefont{Donley et~al.}(2002)\citenamefont{Donley, Claussen,
  Thompson, and Weiman}}]{donley2}
\bibinfo{author}{\bibfnamefont{E.~A.} \bibnamefont{Donley}},
  \bibinfo{author}{\bibfnamefont{N.~R.} \bibnamefont{Claussen}},
  \bibinfo{author}{\bibfnamefont{S.~T.} \bibnamefont{Thompson}},
  \bibnamefont{and} \bibinfo{author}{\bibfnamefont{C.~E.}
  \bibnamefont{Weiman}}, \bibinfo{journal}{Nature}
  \textbf{\bibinfo{volume}{417}}, \bibinfo{pages}{529} (\bibinfo{year}{2002}).

\bibitem[{\citenamefont{Duine and Stoof}(2004)}]{feshreview}
\bibinfo{author}{\bibfnamefont{R.~A.} \bibnamefont{Duine}} \bibnamefont{and}
  \bibinfo{author}{\bibfnamefont{H.~T.~C.} \bibnamefont{Stoof}},
  \bibinfo{journal}{Physics Reports} \textbf{\bibinfo{volume}{396}},
  \bibinfo{pages}{115} (\bibinfo{year}{2004}).

\bibitem[{\citenamefont{Kleppner}(2004)}]{klep}
\bibinfo{author}{\bibfnamefont{D.}~\bibnamefont{Kleppner}},
  \bibinfo{journal}{Physics Today} \textbf{\bibinfo{volume}{57}}, \bibinfo{number}{No. 8}, \bibinfo{pages}{12} (\bibinfo{year}{2004}).

\bibitem[{\citenamefont{Regal et~al.}(2003)\citenamefont{Regal, Ticknor, Bohn,
  and Jin}}]{regal1}
\bibinfo{author}{\bibfnamefont{C.~A.} \bibnamefont{Regal}},
  \bibinfo{author}{\bibfnamefont{C.}~\bibnamefont{Ticknor}},
  \bibinfo{author}{\bibfnamefont{J.~L.} \bibnamefont{Bohn}}, \bibnamefont{and}
  \bibinfo{author}{\bibfnamefont{D.~S.} \bibnamefont{Jin}},
  \bibinfo{journal}{Nature} \textbf{\bibinfo{volume}{424}}, \bibinfo{pages}{47}
  (\bibinfo{year}{2003}).

\bibitem[{\citenamefont{Romans et~al.}(2004)\citenamefont{Romans, Duine,
  Sachdev, and Stoof}}]{romans}
\bibinfo{author}{\bibfnamefont{M.~W.~J.} \bibnamefont{Romans}},
  \bibinfo{author}{\bibfnamefont{R.~A.} \bibnamefont{Duine}},
  \bibinfo{author}{\bibfnamefont{S.}~\bibnamefont{Sachdev}}, \bibnamefont{and}
  \bibinfo{author}{\bibfnamefont{H.~T.~C.} \bibnamefont{Stoof}},
  \bibinfo{journal}{Physical Review Letters} \textbf{\bibinfo{volume}{93}},
  \bibinfo{pages}{020405} (\bibinfo{year}{2004}).

\bibitem[{\citenamefont{Dickerscheid et~al.}(2005)\citenamefont{Dickerscheid,
  Khawaja, Osten, and Stoof}}]{stoof2}
\bibinfo{author}{\bibfnamefont{D.~B.~M.} \bibnamefont{Dickerscheid}},
  \bibinfo{author}{\bibfnamefont{A.}~\bibnamefont{Khawaja}},
  \bibinfo{author}{\bibfnamefont{D.~van} \bibnamefont{Osten}}, \bibnamefont{and}
  \bibinfo{author}{\bibfnamefont{H.~T.~C.} \bibnamefont{Stoof}},
  \bibinfo{journal}{Physical Review A} \textbf{\bibinfo{volume}{71}}, \bibinfo{pages}{043604}
  (\bibinfo{year}{2005}).

\bibitem[{\citenamefont{Kurtzke}(1993)}]{kurtzke}
\bibinfo{author}{\bibfnamefont{C.}~\bibnamefont{Kurtzke}},
  \bibinfo{journal}{IEEE Photonics Technology Letters}
  \textbf{\bibinfo{volume}{5}}, \bibinfo{pages}{1250} (\bibinfo{year}{1993}).

\bibitem[{\citenamefont{Kevrekidis
  et~al.}(2003{\natexlab{b}})\citenamefont{Kevrekidis, Theocharis,
  Frantzeskakis, and Malomed}}]{FRM}
\bibinfo{author}{\bibfnamefont{P.~G.} \bibnamefont{Kevrekidis}},
  \bibinfo{author}{\bibfnamefont{G.}~\bibnamefont{Theocharis}},
  \bibinfo{author}{\bibfnamefont{D.~J.} \bibnamefont{Frantzeskakis}},
  \bibnamefont{and} \bibinfo{author}{\bibfnamefont{B.~A.}
  \bibnamefont{Malomed}}, \bibinfo{journal}{Physical Review Letters}
  \textbf{\bibinfo{volume}{90}}, \bibinfo{pages}{230401}
  (\bibinfo{year}{2003}{\natexlab{b}}).

\bibitem[{\citenamefont{Abdullaev et~al.}(2003)\citenamefont{Abdullaev, Tsoy,
  Malomed, and Kraenkel}}]{tsoy}
\bibinfo{author}{\bibfnamefont{F.~K.} \bibnamefont{Abdullaev}},
  \bibinfo{author}{\bibfnamefont{E.~N.} \bibnamefont{Tsoy}},
  \bibinfo{author}{\bibfnamefont{B.~A.} \bibnamefont{Malomed}},
  \bibnamefont{and} \bibinfo{author}{\bibfnamefont{R.~A.}
  \bibnamefont{Kraenkel}}, \bibinfo{journal}{Physical Review A}
  \textbf{\bibinfo{volume}{68}}, \bibinfo{pages}{053606} (\bibinfo{year}{2003}).

\bibitem[{\citenamefont{Brazhnyi and Konotop}(2005)}]{feshol}
\bibinfo{author}{\bibfnamefont{V.~A.} \bibnamefont{Brazhnyi}} \bibnamefont{and}
  \bibinfo{author}{\bibfnamefont{V.~V.} \bibnamefont{Konotop}}, \bibinfo{journal}{Physical Review A}
  \textbf{\bibinfo{volume}{72}}, \bibinfo{pages}{033615} (\bibinfo{year}{2005}).

\bibitem[{\citenamefont{Matuszewski et~al.}(To
  appear)\citenamefont{Matuszewski, Infeld, Malomed, and Trippenbach}}]{matu2}
\bibinfo{author}{\bibfnamefont{M.}~\bibnamefont{Matuszewski}},
  \bibinfo{author}{\bibfnamefont{E.}~\bibnamefont{Infeld}},
  \bibinfo{author}{\bibfnamefont{B.~A.} \bibnamefont{Malomed}},
  \bibnamefont{and}
  \bibinfo{author}{\bibfnamefont{M.}~\bibnamefont{Trippenbach}},
  \bibinfo{journal}{Physical Review Letters} \textbf{\bibinfo{volume}{95}}, \bibinfo{pages}{050403}
  (\bibinfo{year}{2005}).

\bibitem[{\citenamefont{K\"ohl et~al.}(2005)\citenamefont{K\"ohl, Moritz,
  St\"oferle, G\"unter, and Esslinger}}]{esslinger}
\bibinfo{author}{\bibfnamefont{M.}~\bibnamefont{K\"ohl}},
  \bibinfo{author}{\bibfnamefont{H.}~\bibnamefont{Moritz}},
  \bibinfo{author}{\bibfnamefont{T.}~\bibnamefont{St\"oferle}},
  \bibinfo{author}{\bibfnamefont{K.}~\bibnamefont{G\"unter}}, \bibnamefont{and}
  \bibinfo{author}{\bibfnamefont{T.}~\bibnamefont{Esslinger}},
  \bibinfo{journal}{Physical Review Letters} \textbf{\bibinfo{volume}{94}}, \bibinfo{pages}{080403}
  (\bibinfo{year}{2005}).

\bibitem[{\citenamefont{Pelinovsky
  et~al.}(2004{\natexlab{a}})\citenamefont{Pelinovsky, Kevrekidis,
  Frantzeskakis, and Zharnitsky}}]{hamavg}
\bibinfo{author}{\bibfnamefont{D.~E.} \bibnamefont{Pelinovsky}},
  \bibinfo{author}{\bibfnamefont{P.~G.} \bibnamefont{Kevrekidis}},
  \bibinfo{author}{\bibfnamefont{D.~J.} \bibnamefont{Frantzeskakis}},
  \bibnamefont{and}
  \bibinfo{author}{\bibfnamefont{V.}~\bibnamefont{Zharnitsky}},
  \bibinfo{journal}{Physical Review E} \textbf{\bibinfo{volume}{70}}, \bibinfo{pages}{047604}
  (\bibinfo{year}{2004}{\natexlab{a}}).

\bibitem[{\citenamefont{Zharnitsky and Pelinovsky}(2005)}]{zharn}
\bibinfo{author}{\bibfnamefont{V.}~\bibnamefont{Zharnitsky}} \bibnamefont{and}
  \bibinfo{author}{\bibfnamefont{D.~E.} \bibnamefont{Pelinovsky}},
  \bibinfo{journal}{Chaos}  \textbf{\bibinfo{volume}{15}},
  \bibinfo{pages}{037105} (\bibinfo{year}{2005}) 
  
\bibitem[{\citenamefont{Pelinovsky
  et~al.}(2004{\natexlab{b}})\citenamefont{Pelinovsky, Sukhorukov, and
  Kivshar}}]{pelsukhkiv}
\bibinfo{author}{\bibfnamefont{D.~E.} \bibnamefont{Pelinovsky}},
  \bibinfo{author}{\bibfnamefont{A.~A.} \bibnamefont{Sukhorukov}},
  \bibnamefont{and} \bibinfo{author}{\bibfnamefont{Y.~S.}
  \bibnamefont{Kivshar}}, \bibinfo{journal}{Physical Review E}
  \textbf{\bibinfo{volume}{70}}, \bibinfo{pages}{036618}
  (\bibinfo{year}{2004}{\natexlab{b}}).




\bibitem[{\citenamefont{Denschlag
  et~al.}(2004{\natexlab{b}})\citenamefont{Denschlag, Simsarian, H\"{a}ffner, McKenzie, Browaeys, Cho, Helmerson, Rolston, and Phillips}}]{olstandard}
\bibinfo{author}{\bibfnamefont{J. Hecker} \bibnamefont{Denschlag}},
\bibinfo{author}{\bibfnamefont{J.~E.} \bibnamefont{Simsarian}},
\bibinfo{author}{\bibfnamefont{H.} \bibnamefont{Haffner}},
\bibinfo{author}{\bibfnamefont{C.} \bibnamefont{McKenzie}},
\bibinfo{author}{\bibfnamefont{A.} \bibnamefont{Browaeys}},
\bibinfo{author}{\bibfnamefont{D.} \bibnamefont{Cho}},
\bibinfo{author}{\bibfnamefont{K.} \bibnamefont{Helmerson}},
\bibinfo{author}{\bibfnamefont{S.~L.} \bibnamefont{Rolston}}, \bibnamefont{and}
\bibinfo{author}{\bibfnamefont{W.~D.} \bibnamefont{Phillips}},
  \bibinfo{journal}{Journal of Physics B: Atomic Molecular and Optical Physics}
  \textbf{\bibinfo{volume}{35}}, \bibinfo{pages}{3095-3110}
  (\bibinfo{year}{2002}{\natexlab{b}}).

\bibitem[{\citenamefont{Cubizolles et~al.}(2003)\citenamefont{Cubizolles, Bourdel,
  Kokkelmans, Shlyapnikov, and Salomon}}]{cubi}
\bibinfo{author}{\bibfnamefont{J.}~\bibnamefont{Cubizolles}},
  \bibinfo{author}{\bibfnamefont{T.}~\bibnamefont{Bourdel}},
  \bibinfo{author}{\bibfnamefont{S.~J.~J.~M.~F.}~\bibnamefont{Kokkelmans}}, 
  \bibinfo{author}{\bibfnamefont{G.~V.}~\bibnamefont{Shlyapnikov}}, \bibnamefont{and}
  \bibinfo{author}{\bibfnamefont{C.}~\bibnamefont{Salomon}},
  \bibinfo{journal}{Physical Review Letters} \textbf{\bibinfo{volume}{91}},  \bibinfo{pages}{240401} (\bibinfo{year}{2003}).




\bibitem[{\citenamefont{Agueev and Pelinovsky}(1994)}]{agueev}
\bibinfo{author}{\bibfnamefont{D.} \bibnamefont{Agueev}}
  \bibnamefont{and} \bibinfo{author}{\bibfnamefont{D.} \bibnamefont{Pelinovsky}},
  \bibinfo{journal}{SIAM Journal of Applied Mathematics} \textbf{\bibinfo{volume}{65}},
  \bibinfo{pages}{1101} (\bibinfo{year}{2005}).
    
\bibitem[{\citenamefont{Chugunova and Pelinovsky}(2006)}]{chugpel}
\bibinfo{author}{\bibfnamefont{M.}~\bibnamefont{Chugunova}} \bibnamefont{and}
  \bibinfo{author}{\bibfnamefont{D.~E.}~\bibnamefont{Pelinovsky}},
  \bibinfo{journal}{SIAM Journal of Applied Dynamical Systems} \textbf{\bibinfo{volume}{5}},
  \bibinfo{pages}{66} (\bibinfo{year}{2006}).

\bibitem[{\citenamefont{Kevrekidis et~al.}(2005)\citenamefont{Kevrekidis, Stefanov, and 
Pelinovsky}}]{kevstefpel}
\bibinfo{author}{\bibfnamefont{P.~G.} \bibnamefont{Kevrekidis}},
  \bibinfo{author}{\bibfnamefont{D.~E.}~\bibnamefont{Pelinovsky}}, \bibnamefont{and}
  \bibinfo{author}{\bibfnamefont{A.}~\bibnamefont{Stefanov}},
  \bibinfo{journal}{Journal of Physics A: Mathematical and General} \textbf{\bibinfo{volume}{39}},
  \bibinfo{pages}{479} (\bibinfo{year}{2006}).

\bibitem[{\citenamefont{Gubeskys et~al.}(2005)\citenamefont{Gubeskys,
  Malomed, and Merhasin}}]{malfesh05}
\bibinfo{author}{\bibfnamefont{A.} \bibnamefont{Gubeskys}},
  \bibinfo{author}{\bibfnamefont{B.~A.} \bibnamefont{Malomed}},
  \bibnamefont{and} \bibinfo{author}{\bibfnamefont{I.~M.}
  \bibnamefont{Merhasin}}, \bibinfo{journal}{Studies in Applied Mathematics}
  \textbf{\bibinfo{volume}{115}}, \bibinfo{pages}{255} (\bibinfo{year}{2005}).

\bibitem[{\citenamefont{Pelinovsky et~al.}(1998)\citenamefont{Pelinovsky,
  Kivshar, and Afanasjev}}]{pelkivafan}
\bibinfo{author}{\bibfnamefont{D.~E.} \bibnamefont{Pelinovsky}},
  \bibinfo{author}{\bibfnamefont{Y.~S.} \bibnamefont{Kivshar}},
  \bibnamefont{and} \bibinfo{author}{\bibfnamefont{V.~V.}
  \bibnamefont{Afanasjev}}, \bibinfo{journal}{Physica D}
  \textbf{\bibinfo{volume}{116}}, \bibinfo{pages}{121} (\bibinfo{year}{1998}).

\bibitem[{\citenamefont{de~Sterke and Sipe}(1994)}]{desterkesipe}
\bibinfo{author}{\bibfnamefont{C.~M.} \bibnamefont{de~Sterke}}
  \bibnamefont{and} \bibinfo{author}{\bibfnamefont{J.~E.} \bibnamefont{Sipe}},
  \bibinfo{journal}{Progress in Optics} \textbf{\bibinfo{volume}{33}},
  \bibinfo{pages}{203} (\bibinfo{year}{1994}).

\end{thebibliography}
\end{document}